\documentstyle[aps,epsf,rotate,subfigure]{revtex}

\begin{document}
\draft
\title{Thermostating by Deterministic Scattering: \\Heat and Shear Flow}
\author{C. Wagner, R. Klages, G. Nicolis}
\address{Center for Nonlinear Phenomena and Complex Systems, 
Universit\'{e} Libre de Bruxelles, Campus Plaine CP 231, Blvd du
Triomphe, B-1050 Brussels, Belgium}
\date{\today}
\maketitle
\begin{abstract} 
We apply a recently proposed novel thermostating mechanism to an
interacting many-particle system where the bulk particles are moving
according to Hamiltonian dynamics. At the boundaries the system is
thermalized by deterministic and time-reversible scattering. We show
how this scattering mechanism can be related to stochastic boundary
conditions. We subsequently simulate nonequilibrium steady states
associated to thermal conduction and shear flow for a hard disk
fluid. The bulk behavior of the model is studied by comparing the
transport coefficients obtained from computer simulations to
theoretical results. Furthermore, thermodynamic entropy production and
exponential phase-space contraction rates in the stationary
nonequilibrium states are calculated showing that in general these
quantities do not agree.
\end{abstract}
\section{Introduction}
Driving macroscopic systems out of equilibrium requires external
forces. Now, the very existence of a nonequilibrium steady state
implies that the temperature of the system must remain time
independent. One way to prevent the system from heating up
indefinitely in nonequilibrium is the introduction of a thermostating
algorithm \cite{AT87}. Starting from molecular dynamics simulations
Evans, Hoover, Nos\'e and others proposed deterministic thermostats to
model equilibrium and nonequilibrium fluids
\cite{Ev83,HLM82,EH83,Nose84a,Nose84b,Hoov85}. In this formalism the
(average) internal energy of the dynamical system is kept constant by
subjecting the particles to fictitious frictional forces, thus leading
to microcanonical or canonical distributions in phase space
\cite{EvMo90,Hoov91,Hess96,MoDe98}. The major feature of this
mechanism is its deterministic and time-reversible character, which is
in contrast to stochastic thermostats
\cite{AT87,LeSp78,TCG82,GKI85,ChLe95,ChLe97}. This allows to
elaborate on  the connection between microscopic
reversibility and macroscopic irreversibility and has led to interesting
new links between statistical physics and dynamical systems theory
\cite{EvMo90,Hoov91,MoDe98,DoVL}, especially to relations between
transport coefficients and Lyapunov exponents
\cite{MH87,PoHo88,ECM,Vanc,BarEC}, and between  entropy
production and phase-space contraction
\cite{HHP87,PoHo87,PoHo88,ChLe95,ChLe97,Rue96,Rue97,VTB97,BTV98}.
Although used almost exclusively in the context of nonequilibrium
systems, the abovementioned thermostating mechanism presents the
drawback that the dynamical equations themselves are altered, even in
equilibrium. This raises the question of whether some of the results
are due to the special nature of this thermostating formalism or are
of general validity
\cite{ChLe95,ChLe97,MaHo92,NiDa98,Mare97,TGN98,Gasp}.

Recently, an alternative mechanism in which thermalization is achieved
in a deterministic and time-reversible way has been put forward by
Klages et al.\ \cite{KRN99,RKN98} and has been applied to a periodic
Lorentz gas under an external field. In the present paper we apply
this thermostating method to an interacting many-particle system
subjected to nonequilibrium boundary conditions giving rise to thermal
conduction and to shear flow. The model is closely related to that of
Chernov and Lebowitz \cite{ChLe95,ChLe97,DePo97b,BCL98}, who study a
hard disk fluid driven out of equilibrium into a steady state shear
flow by applying special scattering rules at the boundaries in which
the particle velocity is kept constant. Our model is introduced in
Section
\ref{sec1} where the thermalization mechanism is tested under
equilibrium conditions. In Section \ref{sec2} we move on to the case
of an imposed temperature gradient and a velocity field by adapting
the scattering rules, and we compute the respective transport
coefficients. Having a deterministic and time-reversible system at
hand we proceed in Section \ref{sec3} to investigate the relation
between thermodynamic entropy production and phase-space contraction
rate in nonequilibrium stationary states. The main conclusions are
drawn in Section \ref{sec4}.
\section{Equilibrium state}\label{sec1}
Consider a two-dimensional system of hard discs confined in a square
box of length $L$ with periodic boundary conditions along the x-axis,
i.e., the left and right sides at $x=\pm L/2$ are identified. At the
top and bottom sides of the box, $y=\pm L/2$, we introduce rigid walls
where the discs are reflected according to certain rules to be defined
later. The discs interact among themselves via impulsive hard
collisions so that the bulk dynamics is purely conservative. In the
following and in all the numerical computations we use reduced units
by setting the particle mass $m$, the disk diameter $\sigma$ and the
Boltzmann constant $k_B$ equal to one.

Before proceeding to the nonequilibrium case we define the disc-wall
collision rules in equilibrium and check whether the system is
well-behaved. Now, in equilibrium the bulk distribution is  Gaussian with a
temperature $T$, and the in- and outgoing fluxes at the top and bottom
wall  have the form (see \cite{LeSp78,TCG82,GKI85,ChLe95,ChLe97})
\begin{equation}\label{e1}
\Phi(v_x,v_y)=(2\pi T^3)^{-1/2}|v_y|\exp\left(-\frac{v_x^2+v_y^2}{2T}\right), 
\end{equation} 
with $v_y<0$ for the bottom wall and $v_y>0$ for the top
wall. Imposing stochastic boundary conditions on the systems in this setting would
mean that for every incoming particle the outgoing velocities are
chosen randomly according to Eq. (\ref{e1}). In practice, this is
usually done by drawing numbers from two independent uniformly
distributed random generators $\zeta ,\xi \in[0,1]$ and then transforming these
numbers with the invertible map $\mbox{\boldmath${\cal T}$ }^{-1}:[0,1]\times[0,1]\to
[0,\infty)\times[0,\infty)$ as  
\begin{equation}\label{e2}
(v_x,v_y) = \mbox{\boldmath${\cal T}$ }^{-1} (\zeta,\xi)=\sqrt{2T}\left(\mbox{erf}^{-1}(\zeta),
\sqrt{-\ln(\xi)}\right) ,
\end{equation}
which amounts to transforming the uniform densities $\rho(\zeta)=1$ and
$\rho(\xi)=1$ onto $\Phi(v_x,v_y)$ according to
\begin{equation}\label{e3}  
\rho(\zeta)\rho(\xi)\left|\frac{d\zeta d\xi}{dv_xdv_y}\right|=\left|\frac{\partial \mbox{\boldmath${\cal T}$ }(v_x,v_y)}{\partial v_x\partial v_y}\right|=(2/\pi T^3)^{1/2}|v_y|\exp\left(-\frac{v_x^2+v_y^2}{2T}\right) .
\end{equation}
Note that so far we have restricted Eqs.\ (\ref{e2}),(\ref{e3}) to
positive velocities $v_x,v_y\in[0,\infty)$, which implies a
normalization factor in Eq.\ (\ref{e3}) being different to the one of
Eq.\ (\ref{e1}).  In analogy to stochastic boundaries we now define
the deterministic scattering at the walls as follows. First, take the
incoming velocities $v_x$, $v_y$ and transform them via
$\mbox{\boldmath${\cal T}$ }(v_x,v_y)=(\zeta,\xi)$ onto the unit
square. Second, use a two-dimensional, invertible, phase-space
conserving chaotic map ${\cal M}:[0,1]\times[0,1] \to[0,1]\times[0,1]$
to obtain $(\zeta',\xi')={\cal M}(\zeta,\xi)$. Finally, transform back
to the outgoing velocities via $(v_x',v_y')=\mbox{\boldmath${\cal T}$
}^{-1}(\zeta',\xi')$. In order to render the collision process
time-reversible, we also have to distinguish between particles with
positive and negative tangential velocities by using ${\cal M}$ and
${\cal M}^{-1}$, respectively. Thus, particles going in with positive
(negative) velocities have to go out with positive (negative)
velocities and the full collision rules read
\begin{eqnarray}\label{e4}
(v_x',v_y')&=&\mbox{\boldmath${\cal T}$ }^{-1}\circ{\cal M}\circ\mbox{\boldmath${\cal T}$ }  (v_x,v_y) , \qquad v_x\ge 0 \\
(v_x',v_y')&=&\mbox{\boldmath${\cal T}$ }^{-1}\circ{\cal M}^{-1}\circ\mbox{\boldmath${\cal T}$ }  (v_x,v_y) , \quad v_x<0\nonumber,
\end{eqnarray} 
where $\mbox{\boldmath${\cal T}$ }$ is meant to be applied to the modulus of the velocities
\cite{alter}. Since both the positive and the negative side of the tangential
velocity distribution of Eq.\ (\ref{e1}) are normalized to $1/2$, this
normalization factor has to be incorporated in Eq.(\ref{e3}) to render
the full desired flux $\Phi$ equivalent to the one of Eq.\
(\ref{e1}). Rewriting Eq.\ (\ref{e3}) in polar coordinates yields
precisely the transformation used in
\cite{KRN99,RKN98}, in the limiting case where it mimics a reservoir
with infinitely many degrees of freedom. It should also be realized 
that for obtaining the transformation $\mbox{\boldmath${\cal T}$ }$ the total number of
degrees of freedom of the reservoir has been projected out onto a
single velocity variable, which couples the bulk to the
reservoir. Eckmann et al. \cite{EPRB98} used a similar idea to go from
a Hamiltonian reservoir with infinitely many degrees of freedom to a
reduced description when modeling heat transfer via a finite chain of
nonlinear oscillators. It remains to assign the form of  the chaotic map ${\cal M}$ and we shall first adopt the choice of a baker map, as in
Refs.\ \cite{KRN99,RKN98},
\begin{equation}\label{e5}
(\zeta',\xi')={\cal M}(\zeta,\xi)=\left\{\begin{array}{l@{\quad;\quad}l}(2\zeta,\xi/2)& 0\le \zeta \le 1/2\\(2\zeta-1,(\xi+1)/2)&1/2<\zeta\le 1\end{array}\right. .
\end{equation}
Later on we will investigate the consequences of choosing other
mappings like the standard map (see, e.g.,
\cite{Ott,Meis92}). Since in equilibrium the in- and outgoing fluxes
have the same form as in Eq.(\ref{e1}), the baker map yields a uniform
density and our scattering prescription in Eq.(\ref{e4}) can be viewed
as a deterministic and time-reversible counterpart of stochastic
boundary conditions. ${\cal M}$ being chaotic, the initial and final
momentum and energy of any single particle are certainly different,
but both quantities should be conserved on the average. The latter is
confirmed by numerical experiments in equilibrium, where, as usual in
hard disk simulations, we follow a collision-to-collision approach
\cite{AT87}.  Keeping the volume fraction
occupied by $N=100$ hard disks equal to $\rho=0.1$ sets the length of
the box equal to $L=28.0$. After some transient behavior which depends
on the temperature of the initial configuration the bulk distribution
is Gaussian with zero mean and mean kinetic energy $T/2$ in each directions.
The in- and outgoing fluxes at the walls are correctly equipartioned
with $T$ as well and have the desired form of Eq.(\ref{e1}), so the
system reproduces  the correct statistic properties. We close this section by a  remark on how we measure the temperature of
a flux to or from the boundaries. As the temperature of the tangential
component we use the variance of the velocity distribution,
$T_x:=\left<(v_x-\left<v_x\right>_x)^2\right>_x$, where $<>_x$ denotes an
average over the density $\rho(v_x)$. On the other hand, since in the normal direction we
actually measure a flux, the appropriate prescription to measure the
temperature of this component is
$T_y:=\left[v_y\right]_y/\left[v_y^{-1}\right]_y$, where
$[]_y$ represents an average over the flux $\Phi$ and the denominator
serves as a normalization. The temperatures of the in- and outgoing
fluxes at the wall are then defined as $T_{i/o}:=(T_x+T_y)/2$, and
$T_w:=(T_i+T_o)/2$.
\section{Nonequilibrium steady state}\label{sec2}
\subsection{Heat flow}\label{sec2.1}
\subsubsection{The Model}
In the following, we explicitly indicate the dependence of the
transformation $\mbox{\boldmath${\cal T}$ }$ on the parameter $T$ by
writing $\mbox{\boldmath${\cal T}$ }\!\!_T$.  This immediately
indicates how we may drive our system to thermal nonequilibrium: We
just have to use different values of this parameter for the upper
($T^u$) and the lower wall ($T^d$). We deliberately avoid to use the
word 'temperature' for this parameter, since in contrast to stochastic
boundary conditions we have generally no idea how a different $T$
affects the actual temperature of the wall in the sense of the
definition given above. In a nonequilibrium situation the temperature
of the ingoing flux $\Phi_i$ generally does not match exactly the
parameter $T$. Therefore, we do not transform onto the uniform
invariant density of the baker map anymore, and consequently, the
outgoing flux might have all kinds of shapes or
temperatures. Nevertheless, the hope is that the mapping ${\cal M}$
will be chaotic enough to smooth out most of the differences and to
produce a reasonable outgoing flux $\Phi_o$ such that the system is
correctly thermostated. And this is indeed what we find in the
numerical experiments.
\subsubsection{Numerical Results}
We  set $T^u=2$, $T^d=1$ and $\rho=0.1$ and average over about 40000
particle-particle collisions per particle and about 6000 particle-wall
collisions per particle. We divide the available vertical height $L-1$ into 20 equally
spaced horizontal layers  and calculate the
time averages of the number density $n(y)$ of the particles, the mean
velocities $u_x(y)=\left<v_x\right>$, $u_y(y)=\left<v_y\right>$, and
the variances $\left\langle(v_x-u_x)^2\right\rangle$,
$\left\langle(v_y-u_y)^2\right\rangle$. Furthermore, we record the
time average of the kinetic energy transfer and measure the
temperatures of the in- and outgoing fluxes of both walls as described
in the preceding section. The temperatures at the walls are then
defined as the mean value of the in- and outgoing temperatures,
$T_w^{u/d}:=(T_i^{u/d}+T_o^{u/d})/2$. Time series plots of these
quantities confirm the existence of a  nonequilibrium stationary state
(NSS) induced by the temperature gradient.

\begin{figure}
\epsfxsize=10cm
\centerline{\epsfbox{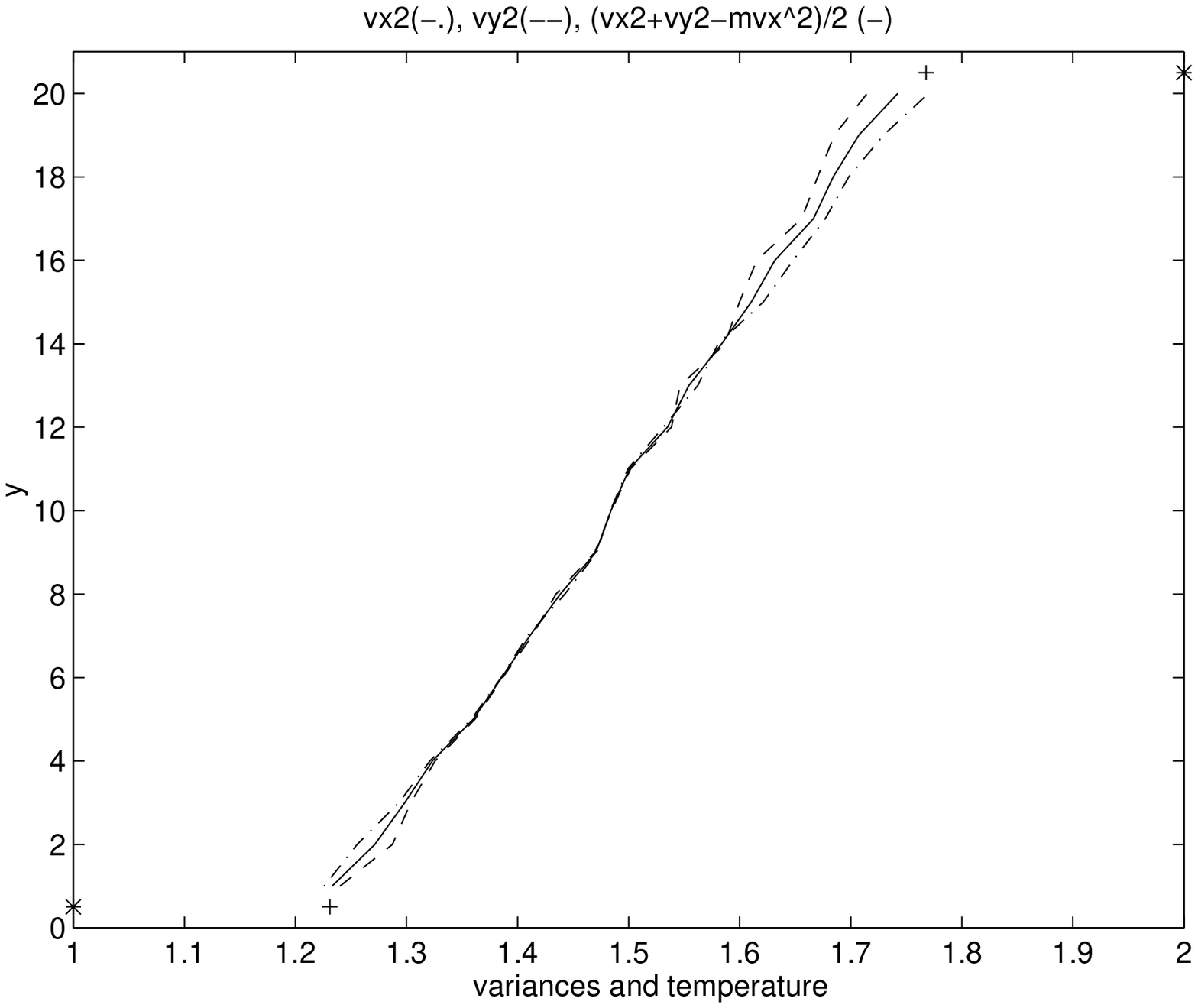}}
%\vspace*{0.3cm} 
\caption{Temperature (-) and variance (- -)/(-.) profiles, $T_u=2$, $T_d=1$ (*). (+) denotes the wall temperatures $T_w$ as defined in the text.}
\label{fig1}
\end{figure}

\begin{figure}
\epsfxsize=10cm
\centerline{\epsfbox{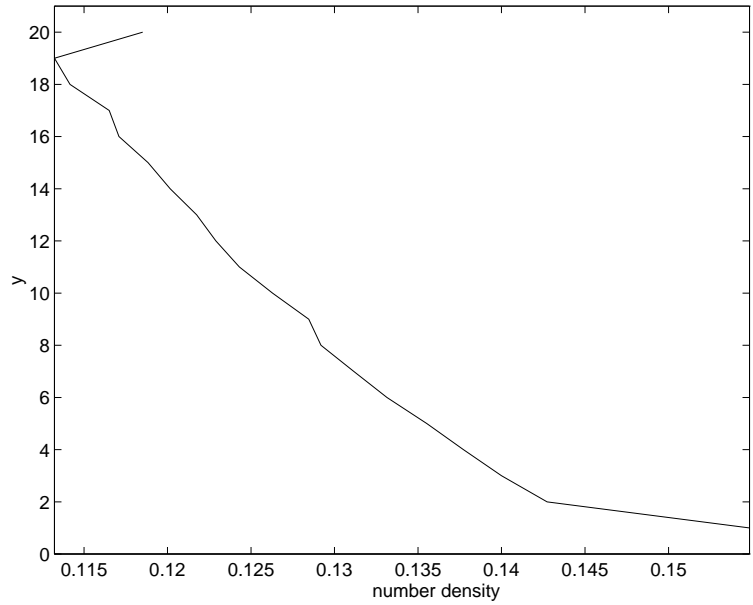}}
%\vspace*{0.3cm} 
\caption{The profile of the number density, $T_u=2$, $T_d=1$ .}
\label{fig2}
\end{figure}

\begin{figure}
\begin{center}
\epsfxsize=6.5cm
\subfigure[]{\epsfbox{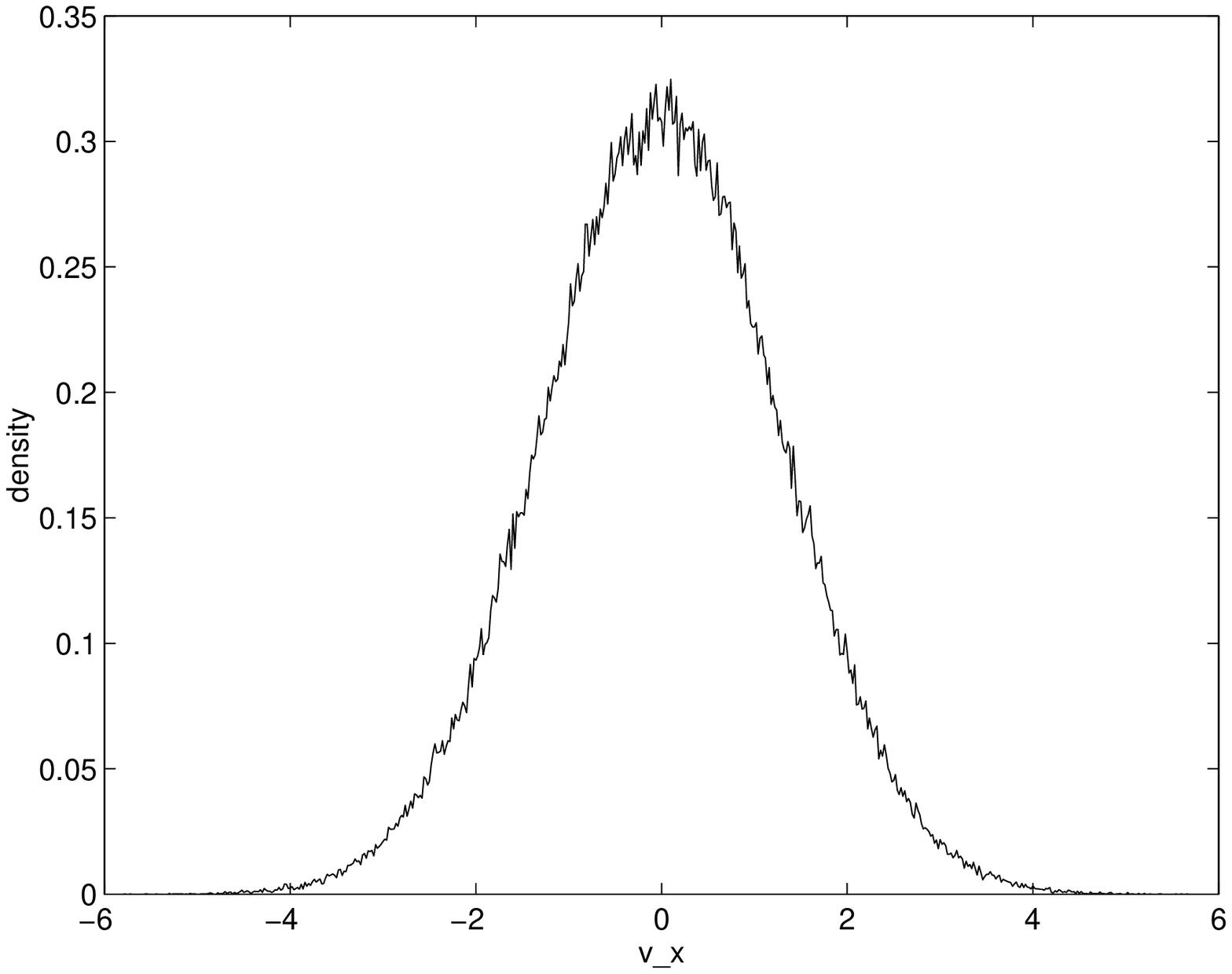}}
\epsfxsize=6.5cm
\subfigure[]{\epsfbox{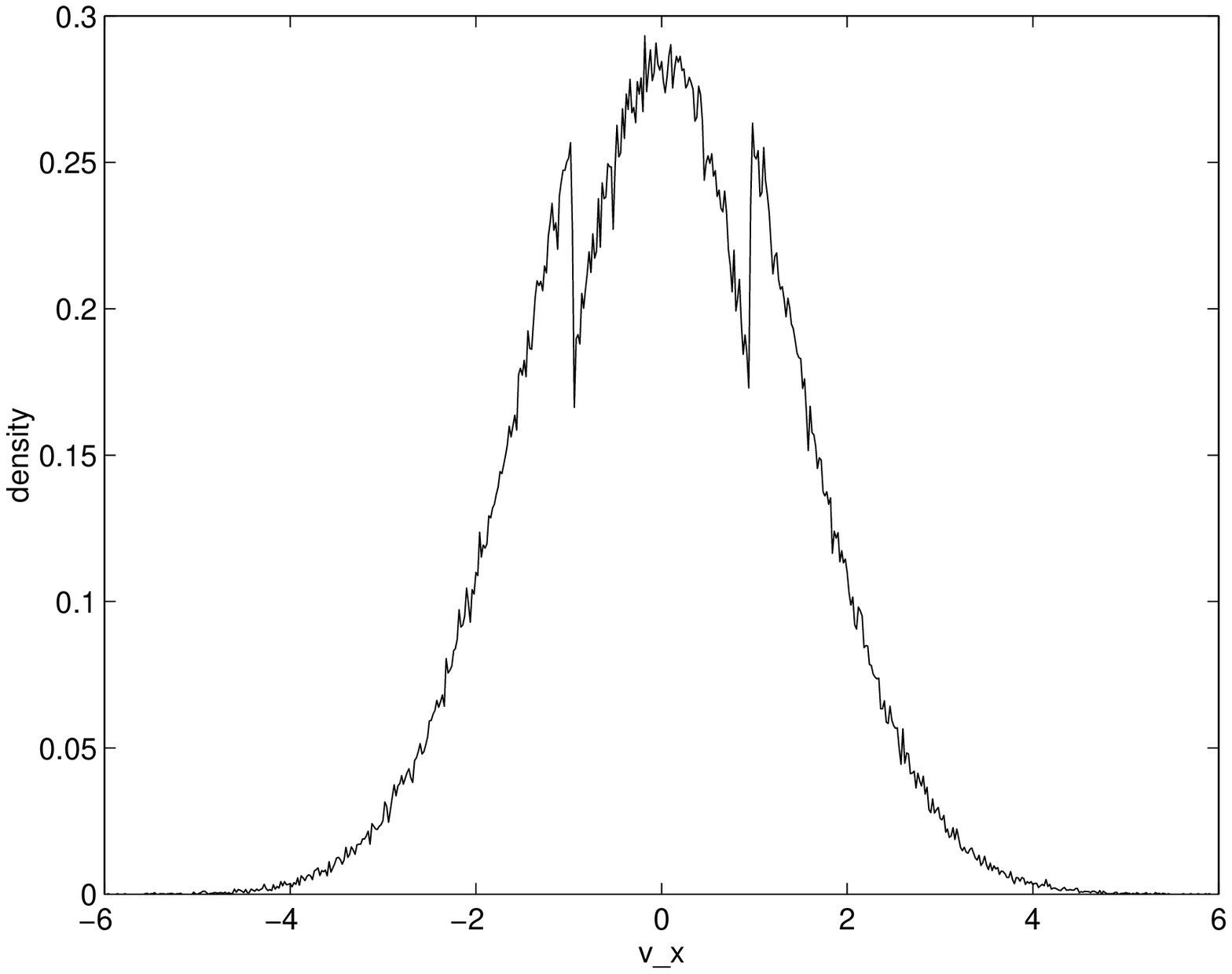}}\\
\epsfxsize=6.5cm
\subfigure[]{\epsfbox{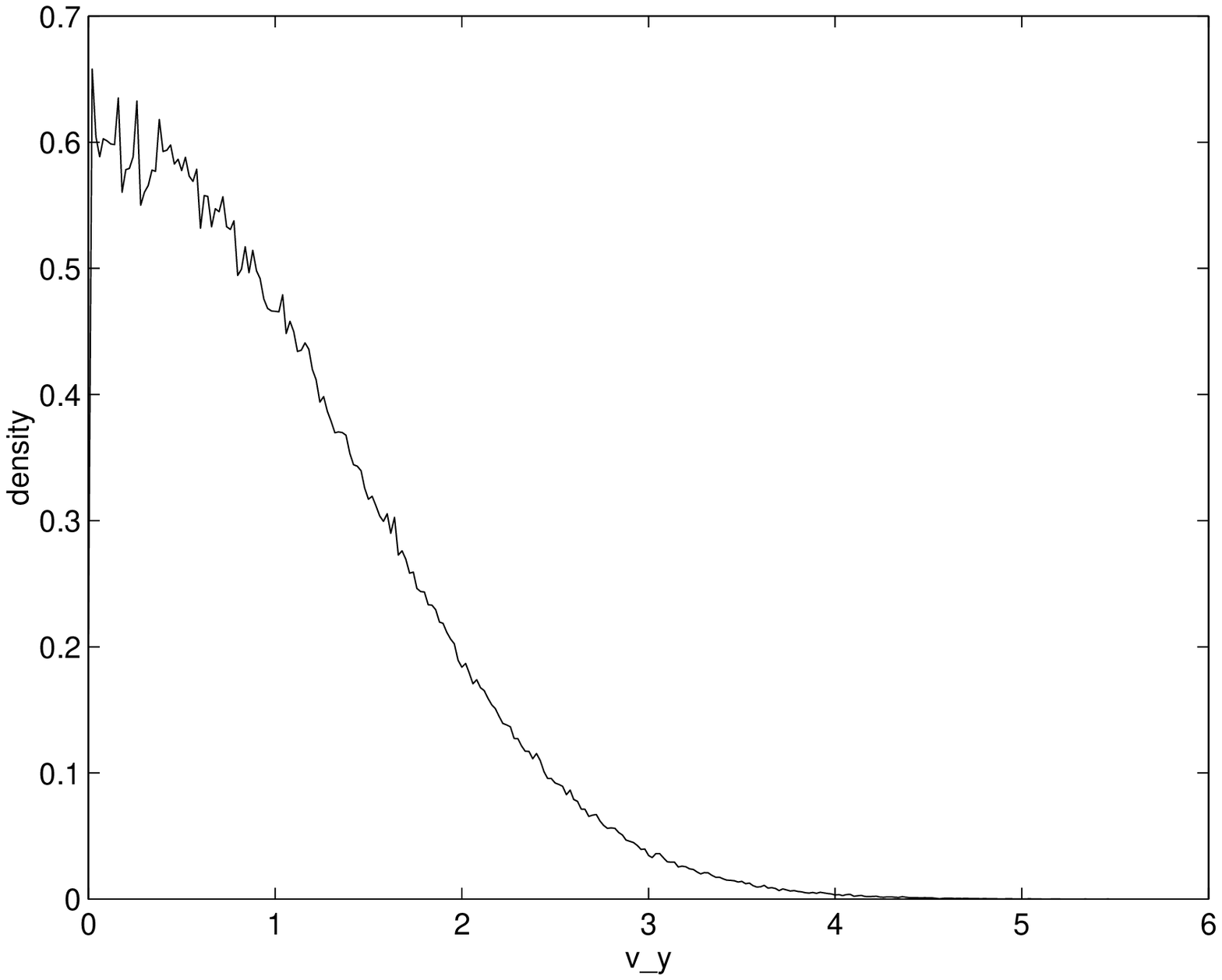}}
\epsfxsize=6.5cm
\subfigure[]{\epsfbox{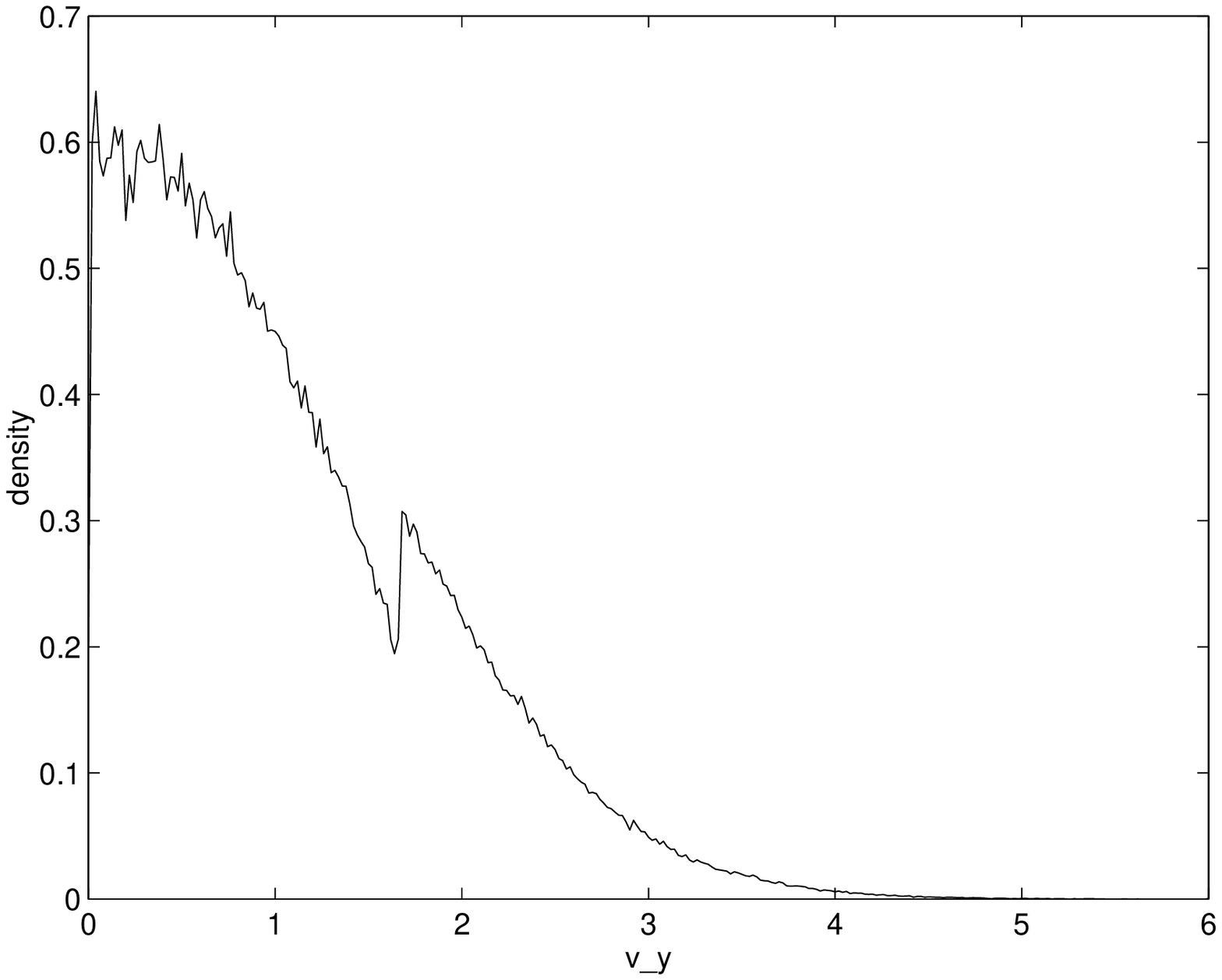}}
\end{center}
\caption{Velocity distributions of the in- and outgoing particles at
the upper wall ($T=2$) for the heat flow case: a) $v_x^{in}$, b)
$v_x^{out}$, c) $v_y^{in}$, d) $v_y^{out}$.}
\label{fig3}
\end{figure}

Fig.\ \ref{fig1} shows the temperature profile between the upper and
the lower wall. Apart from boundary effects it is approximately
linear, and the respective kinetic energy is equipartitioned between
the two degrees of freedom. The parameters $T^u$ and $T^d$ are
represented as (*), and we find a 'temperature' jump, whereas the
measured temperatures $T^{u/d}_w$ (+) at the walls seem to continue
the bulk profile reasonably well. The profile of the number density
$n=4/\pi \rho$ is depicted in Fig.\ \ref{fig2}. Note again the
boundary effects. The densities of the in-coming particles at the
upper wall are Gaussian shaped (Figs. \ref{fig3}a,c), whereas the
outgoing densities (Figs. \ref{fig3}b,d) show cusps due to the folding
property of the baker map. Nevertheless, the baker map produces a
reasonable outgoing flux which generates a NSS.

In order to examine the bulk behavior we now compute the thermal
conductivity in our computer experiment and compare it to the
theoretical value. For this purpose, we measure the heat flux $Q$
across the boundaries and estimate the temperature gradient $dT(y)/dy$
by a linear least square fit to the experimental profile. To discard
boundary effects we use only data in the bulk of the system, namely
from layer 3 to layer 18, i.e.,  excluding the top two and the bottom
two layers. The experimental heat conductivity is then defined as
\begin{equation}\label{e6}
\lambda_{exp}=Q\left(\frac{dy}{dT}\right) ,
\end{equation}
whereas the theoretical expression for the conductivity of a gas of
hard disks with unit mass and unit diameter as predicted by Enskog's
theory reads \cite{RC96,G78}
\begin{equation}\label{e7}
\lambda_l=1.029 2 \sqrt{\frac{T}{\pi}}\left[\frac{1}{\chi}+\frac{3}{2}bn+0.8718(bn)^2\chi\right] .
\end{equation}
Here, $b$ is the second virial coefficient, $b=\pi/2$, and $\chi$ is
the Enskog scaling factor, which is just the pair correlation function
in contact\cite{BH76},
\begin{equation}\label{e8}
\chi=\frac{1-\frac{7}{16}\frac{\pi}{4}n}{(1-\frac{\pi}{4}n)^2}.
\end{equation}
Since (\ref{e7}) depends on local values of $T$ and $n$ we define the
{\it theoretical effective conductivity} $\lambda_{th}$ as the
harmonic mean over the layers \cite{RC96},
\begin{equation}\label{e9}
\lambda_{th}=\left( 1/N_{layers}\sum_{l=1}^{N_{layers}}1/\lambda_l\right)^{-1} .
\end{equation}
Table \ref{tab1} compares $\lambda_{exp}$ to $\lambda_{th}$ by showing
the ratio of the experimental to the theoretical conductivity for
different particle numbers and temperature differences. The agreement
is quite good, so our thermostating mechanism produces a NSS which is
in agreement with hydrodynamics.

Furthermore, going into the hydrodynamic limit by increasing the
number of particles we observe that the discontinuity in the outgoing
flux of Figs. \ref{fig3}(b,d) diminishes,  as expected, since both the
in- and outgoing flux come closer to local equilibrium.
\subsection{Shear Flow}
Inspired by the recently proposed model of Chernov and Lebowitz
\cite{ChLe95,ChLe97} for a boundary driven planar Couette-flow in a
nonequilibrium steady state, we now proceed to check whether it is
possible to combine our thermostating mechanism with a positive
(negative) drift imposed onto the upper (lower) wall,
respectively. Chernov and Lebowitz chose a purely Hamiltonian bulk and
simulated the drift at the boundaries by rotating the angle of the
particle velocity at the moment of the scattering event with the wall
while keeping the absolute value of the velocity constant. This
setting could be formulated in a time-reversible way and keeps the
total energy of the system generically constant. Here we separate the
thermostating mechanism and the drift of the walls by introducing the
map
\begin{equation}\label{e10}
{\cal S}_d(v_x,v_y)=(v_x+d,v_y) ,
\end{equation} 
and by applying this shift to the 'thermostated'
velocities. Time-reversibility forces us to do the same before
thermostating. Thus, the full particle-wall interaction reads
\begin{eqnarray}
\mbox{(Model I)}&&\nonumber\\
(v_x',v_y')&=&{\cal S}_d\circ\mbox{\boldmath${\cal T}$ }\!\!_T^{-1}\circ{\cal
M}\circ\mbox{\boldmath${\cal T}$ }\!\!_T\circ {\cal S}_d (v_x,v_y) , \qquad v_x\ge-d \label{e11}\\
(v_x',v_y')&=&{\cal S}_d\circ\mbox{\boldmath${\cal T}$ }\!\!_T^{-1}\circ{\cal
M}^{-1}\circ\mbox{\boldmath${\cal T}$ }\!\!_T\circ{\cal S}_d (v_x,v_y) , \quad v_x<-d\label{e11b}
,
\end{eqnarray}    
where shifts of different sign are used for the upper (lower) wall to
let the walls move into opposite directions. Other prescriptions to
impose a shear will be investigated in more detail in the following
section.

\begin{figure}
\epsfxsize=10cm
\centerline{\epsfbox{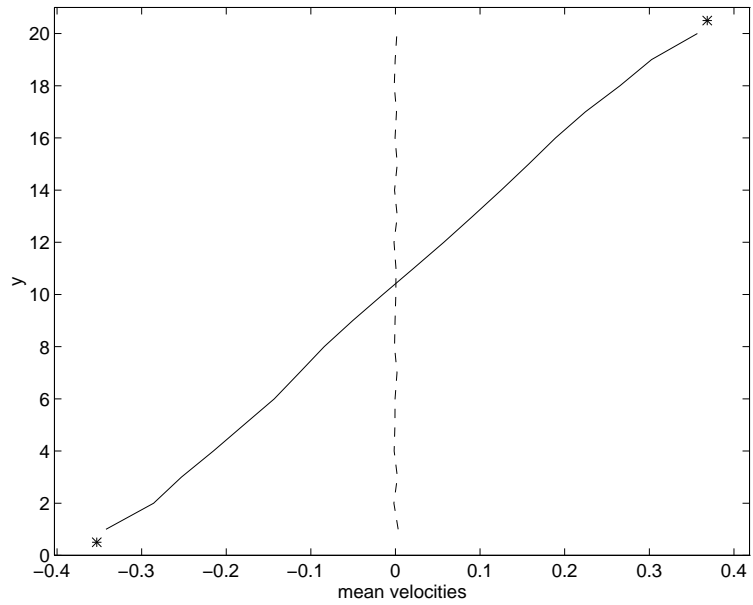}}
%\vspace*{0.3cm} 
\caption{ Mean velocity in the $x$-direction (--), in the $y$-direction (-
-), and at the wall $u_w$ (*).}
\label{fig4}
\end{figure}

\begin{figure}
\epsfxsize=10cm
\centerline{\epsfbox{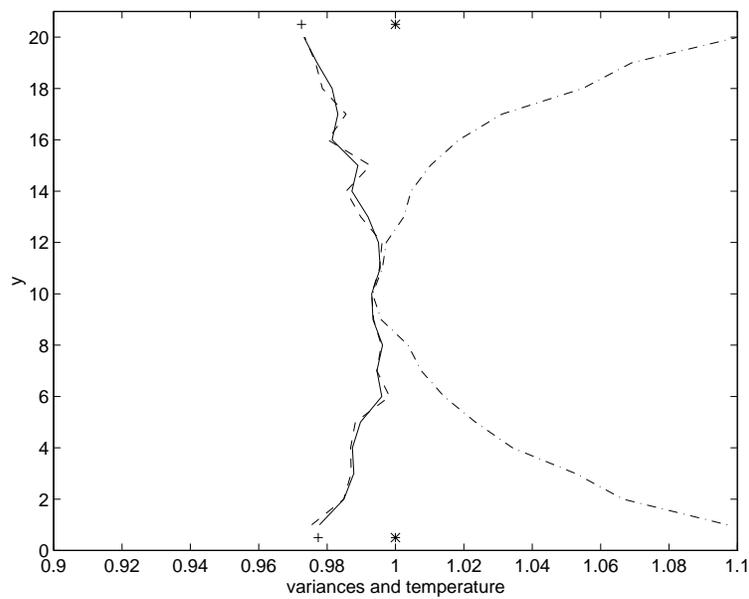}}
%\vspace*{0.3cm} 
\caption{Velocity variance in the $x$-direction (- .) and in the
$y$-direction (- -), temperature (--), measured wall temperature $T_w$
(+) and 'parametrical' temperature $T$ (*).}
\label{fig5}
\end{figure}

In the simulations we set $d=\pm0.05$, $T=T^u=T^d=1.0$, $N=100$ and
$\rho=0.1$. As we expected, we find a NSS with a linear shear profile
along the x-direction (Fig. \ref{fig4}), where the drift velocity of
the wall $u_w$ (*) is defined as the average between the in- and 
outgoing tangential velocities. The temperature profile is shown in
Fig. \ref{fig5}, with the wall temperatures $T_w$ (+) defined as
above. As can be seen in the plots, none of these values correspond to
the parameters $T$ (*) or $d$.  Nevertheless, we obtain a linear shear
profile $u_x(y)$ and an almost quadratic temperature profile $T(y)$,
as predicted by hydrodynamics
\cite{ChLe95}.

\begin{figure}
\begin{center}
\epsfxsize=6.5cm
\subfigure[]{\epsfbox{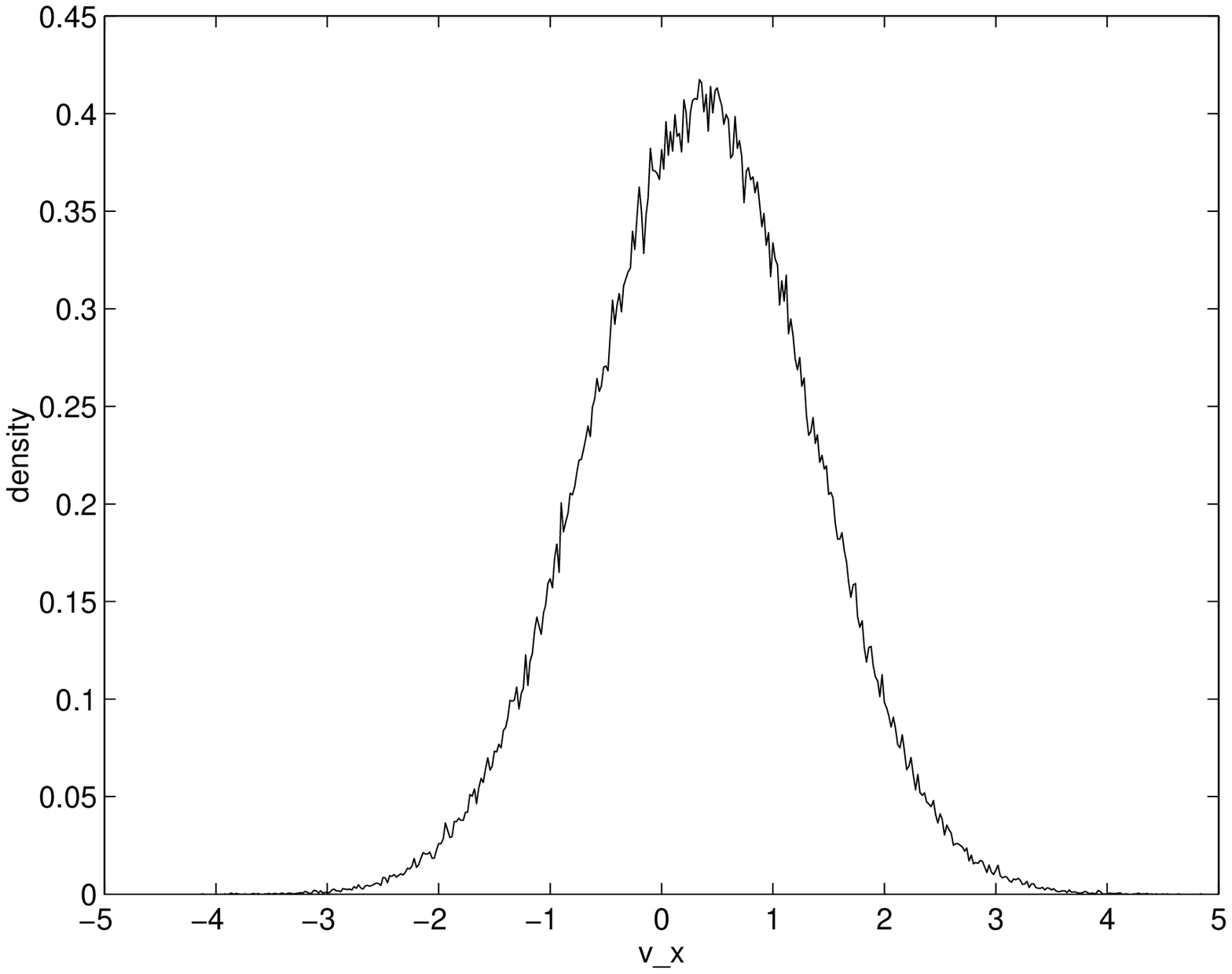}}
\epsfxsize=6.5cm
\subfigure[]{\epsfbox{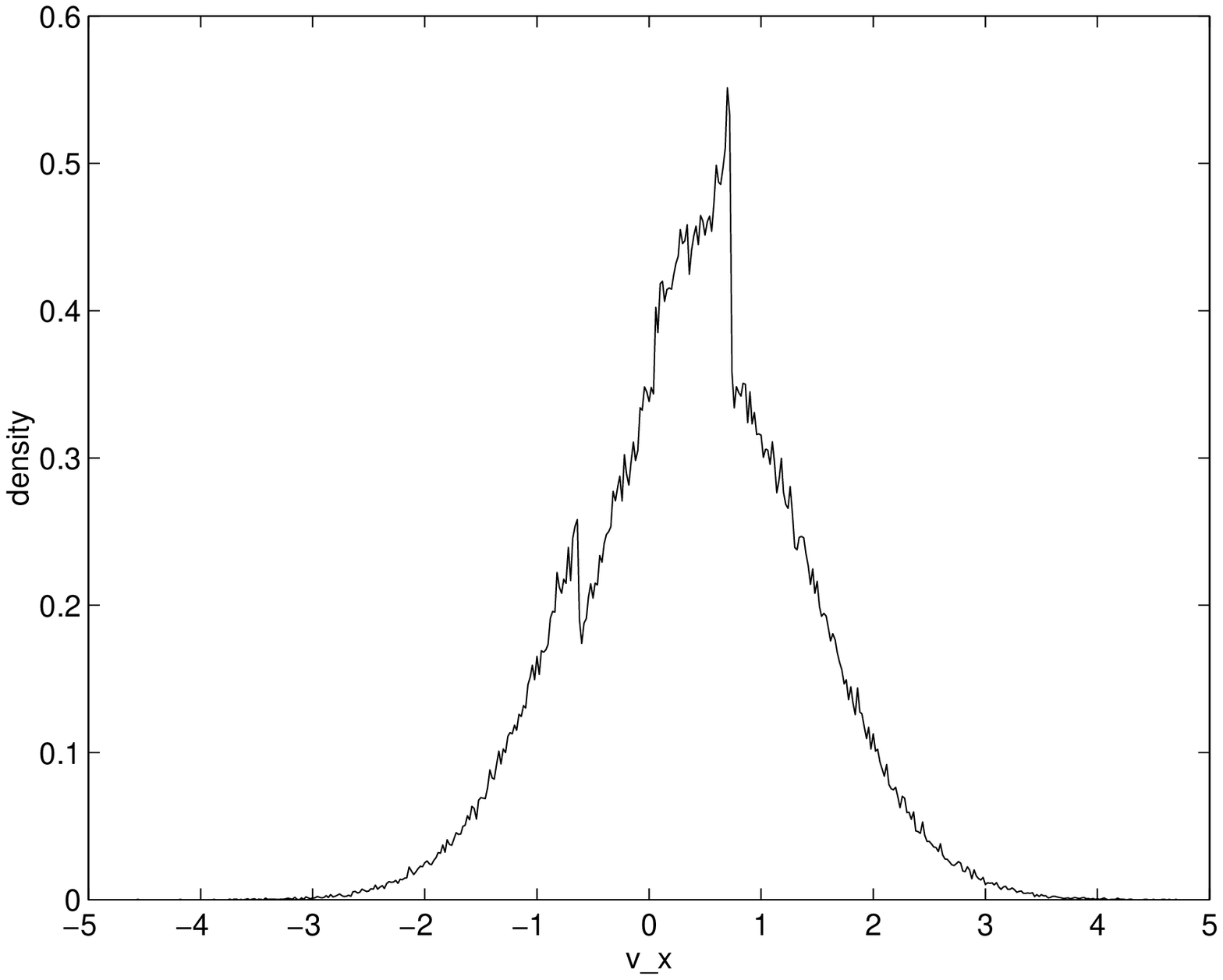}}\\
\epsfxsize=6.5cm
\subfigure[]{\epsfbox{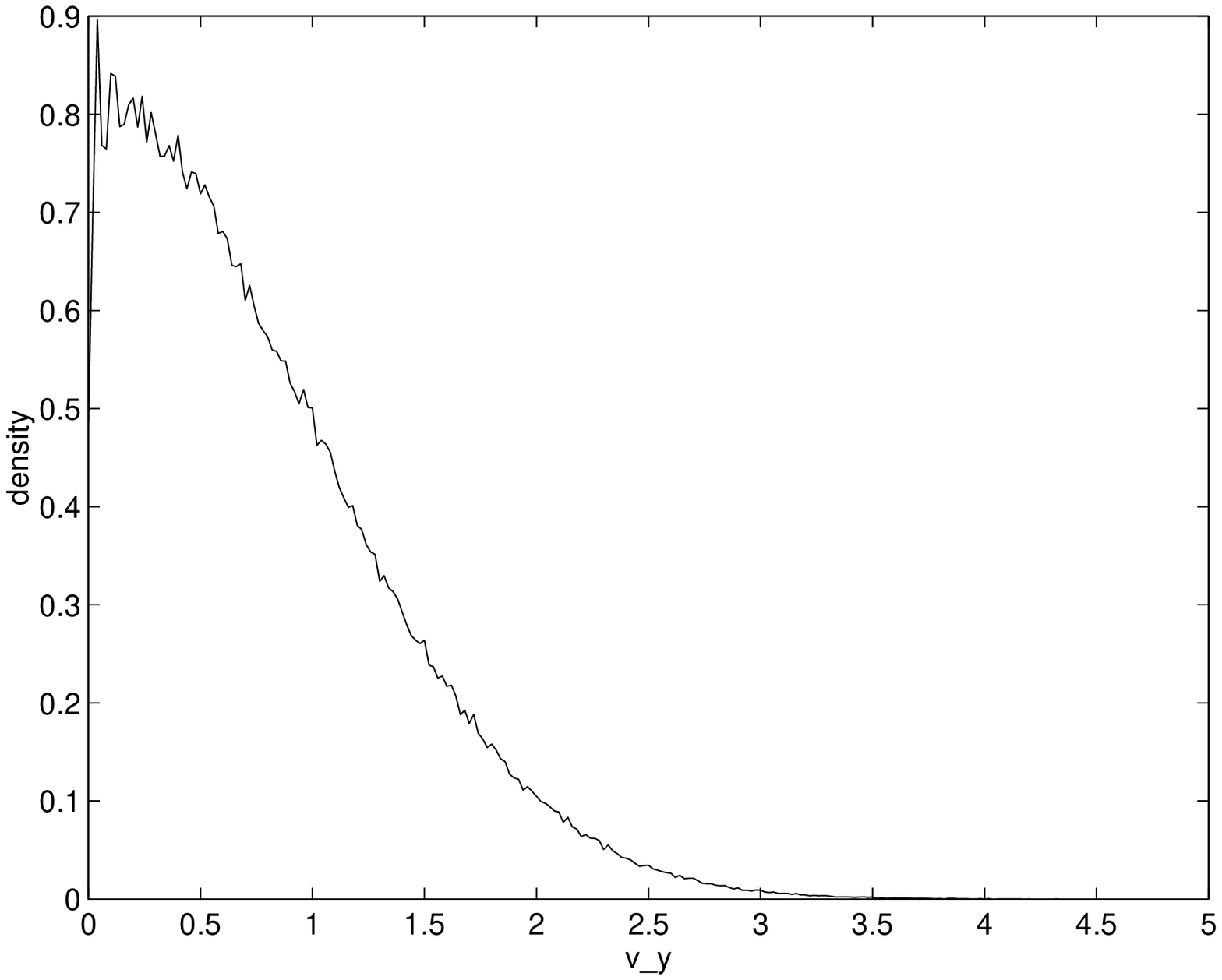}}
\epsfxsize=6.5cm
\subfigure[]{\epsfbox{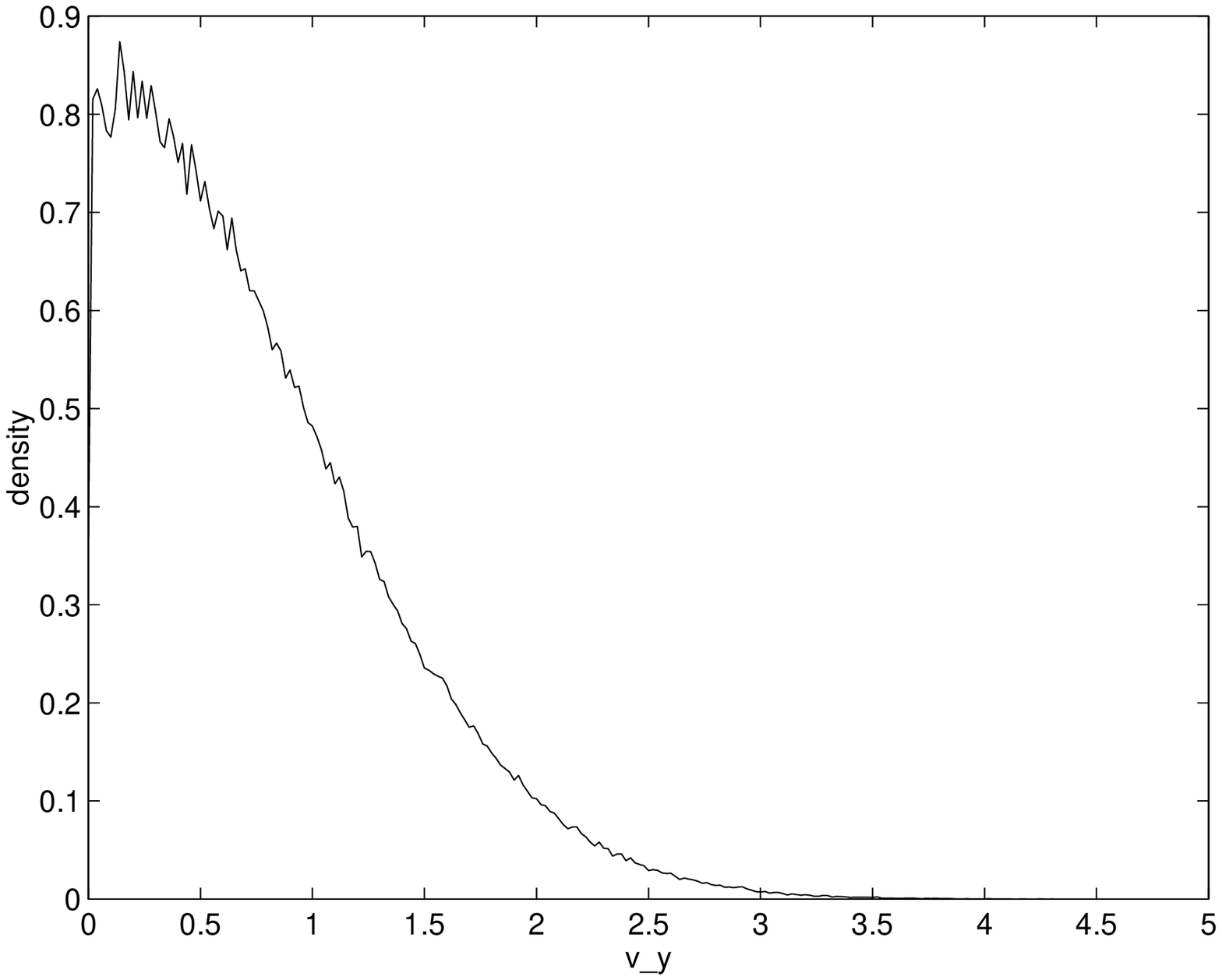}}
\end{center}
\caption{Velocity distributions of the in- and outgoing particles at
the upper wall for the shear flow case (Model I):  a) $v_x^{in}$, b)
$v_x^{out}$, c) $v_y^{in}$, d) $v_y^{out}$.}
\label{fig6}
\end{figure}

For a comparison of experimental and theoretical viscosity we follow
the same procedure as above, i.e., we estimate the experimental shear rate
$du_x(y)/dy$ by a linear least square fit $u_x(y)=\gamma y$, again
discarding the outermost layers. Through the measured momentum
transfer from wall to wall $\Pi$ the experimental viscosity is given
as
\begin{equation}\label{e12}
\eta_{exp}=\Pi/\gamma  ,
\end{equation}
and the theoretical value as calculated by the Enskog theory has the
form \cite{G78,ChLe95}
\begin{equation}\label{e13}
\eta_{l}=1.022\frac{1}{2}
\sqrt{\frac{T}{\pi}}\left[\frac{1}{\chi}+bn+0.8729(bn)^2\chi\right] .
\end{equation}
Again $b=\pi/2$ denotes the second virial coefficient, $\chi$ is given
by Eq.(\ref{e8}) and we use the arithmetic mean of the viscosity over
layers 3 - 18 to compute the theoretical viscosity, i.e.,
$\eta_{th}=1/N_{layers}\sum\eta_{l}$. Table \ref{tab2} shows good
agreement between the experimental and the theoretical values for
different numbers of particles, so again the thermostating mechanism
leads to the correct macroscopic behavior. However, the
discontinuities in the $v_x$-velocity distribution of the scattered
particles should be noticed (see Fig.\ref{fig6}). This time these
discontinuities do not diminish or disappear in the hydrodynamic
limit.

\section{Entropy production and Phase-Space Contraction}\label{sec3}
Having a deterministic and reversible dynamics at hand we can now turn
to properties beyond the usual hydrodynamic ones and, in particular,
investigate the conjectured identity between phase space contraction
rate and thermodynamic entropy production in the light of our
formalism
\cite{HHP87,PoHo87,PoHo88,ChLe95,ChLe97,Rue96,Rue97,VTB97,BTV98}.
In an isolated macroscopic system the entropy is a thermodynamic
potential and therefore plays the central role in determining the time
evolution and the final equilibrium state. Yet, its microscopic
interpretation out of equilibrium remains controversial (see, e.g.,
\cite{Leb93a,Leb93b}) and the situation is even much less clear in
NSS. For the class of models where thermostating is ensured by
friction coefficients an exact equality between entropy production and
phase space contraction rate in NSS has been inferred on the basis of
a global balance between the system and the reservoir
\cite{HHP87,EvMo90,Hoov91,MoDe98}. For the Chernov-Lebowitz model
an approximate equality has also been found
\cite{ChLe95,ChLe97}. Still, it is not clear under which circumstances this
relation holds in general
\cite{ChLe95,ChLe97,MaHo92,NiDa98,Mare97,TGN98,Gasp}. 
\subsection{Equilibrium State}
We begin with the simplest case of equilibrium
 described in section \ref{sec1}, where  the thermodynamic entropy
production ${\overline R}_{eq}$ vanishes. The bulk dynamics being
Hamiltonian,  phase-space contraction can only occur during
collisions with a wall. Since these collisions take place '
instantaneously', we ignore the bulk particles and restrict ourselves
to the compression related to a single collision during the time
interval $dt$. The phase space contraction is then given by the ratio
of the one-particle phase-space volume after the collision $(dx'dy'dv_x'dv_y')$ to
the one before the collision, $(dxdydv_xdv_y)$, and can thus be
obtained from the Jacobi determinant of the scattering process. One easily sees that $dx'=dx$ and
$|dy'/dy|=|v_y'dt/v_ydt|$ \cite{ChLe95,ChLe97}. Furthermore, 
\begin{eqnarray}\label{e14}
\left|\frac{dv_x'dv_y'}{dv_xdv_y}\right|&=&\left| \frac{\partial \mbox{\boldmath${\cal T}$ }}{\partial
v_x \partial v_y}\frac{\partial {\cal M}^{(-1)}}{\partial x \partial y}\frac{\partial
\mbox{\boldmath${\cal T}$ }^{-1}}{\partial v_x' \partial v_y'}  \right|  
=\left| \frac{\partial \mbox{\boldmath${\cal T}$ }}{\partial v_x \partial v_y}\left[\frac{\partial \mbox{\boldmath${\cal T}$ }}{\partial v_x'
\partial v_y'}\right]^{-1} \right|\nonumber\\
&=&\left|\frac{v_y}{v_y'}\right|\exp\left(\frac{v_x'^2+v_y'^2-v_x^2-v_y^2}{2T}\right)
,
\end{eqnarray}
where step two follows from the phase-space conservation of ${\cal
M}/{\cal M}^{-1}$,  and the last line is obtained from
Eqs.(\ref{e2})(\ref{e3}). Hence, in a particle-wall collision the
phase-space volume is changed by a factor of
\begin{equation}\label{e15}
\left|\frac{dv_x'dv_y'dx'dy'}{dv_xdv_ydxdy}\right|=\exp\left(\frac{v_x'^2+v_y'^2-v_x^2-v_y^2}{2T}\right) .
\end{equation}
The mean exponential rate of compression of the phase space volume per
unit time is thus given by
\begin{equation} \label{e16}
\overline P=-<\ln \left|\frac{dv_x'dv_y'dx'dy'}{dv_xdv_ydxdy}\right|>=
\left<(v_x^2+v_y^2-v_x'^2-v_y'^2)/2T\right> ,
\end{equation}
where the brackets $<>$ denote a time average over all collisions at
the top and the bottom walls. In equilibrium the in- and outgoing
fluxes associated to these collisions have the same statistical
properties, so ${\overline P}_{eq}$ sums up to zero and
\begin{equation}\label{e17}
{\overline P}_{eq}={\overline R}_{eq}  ,
\end{equation}
This is fully confirmed by the simulations.
\subsection{NSS}
The thermodynamic entropy production $\sigma$ per unit volume of our
system in NSS is given by the Onsager form \cite{ChLe95,EvMo90}
\begin{equation}\label{e18}
\sigma(y)=\frac{\Pi}{T}\frac{du_x}{dy}+J(y)\frac{d}{dy}\left(\frac{1}{T}\right) ,
\end{equation}
where $\Pi$ is the  x-momentum flux in the negative
y-direction, and $J(y)$ is the heat flux in the positive y-direction.
\subsubsection{Heat Flow}
Imposing only a temperature gradient on our system like in section
\ref{sec2.1}, the first term in Eq.\ (\ref{e18}) is identical to zero and
the total  entropy production $\overline R$ in the steady state
is then 
\begin{equation}\label{e19}
\overline R = \int_{Volume} \!\!\!\sigma d{\bf r}=\int_{Surface}\!\!\!J/T ds= J^u_w/T^u_w+ J^d_w/T^d_w
\end{equation}
The right hand side of Eq.(\ref{e19}) is the outward entropy flux $J_w/T_w$ across the walls of the container. Note that  there is no
temperature slip at the walls with respect to the correctly
defined temperature values, as  indicated by the simulation results
in Figs.\ref{fig1} and \ref{fig5}.

On the other hand, the exponential phase-space contraction rate now reads
\begin{equation} \label{e20}
\overline P=\left<(v_x^2+v_y^2-v_x'^2-v_y'^2)/2T_u\right>_u+\left<(v_x^2+v_y^2-v_x'^2-v_y'^2)/2T_d\right>_d,
\end{equation}
where we  averaged over the upper and the lower wall
separately. Since in NSS
$J_w^u=\left<(v_x^2+v_y^2-v_x'^2-v_y'^2)/2\right>_u=-J_w^d=-\left<(v_x^2+v_y^2-v_x'^2-v_y'^2)/2\right>_d$,
the ratios of entropy production to exponential phase space
contraction rate reduce to
\begin{equation}\label{e21}
\frac{\overline R^{u/d}}{\overline P^{u/d}}=\frac{T^{u/d}}{T^{u/d}_w}.
\end{equation}
In the hydrodynamic limit the in- and the outgoing fluxes approach
local equilibrium, implying $T_i^{u/d}\simeq T_o^{u/d}\simeq
T_w^{u/d}\simeq T^{u/d}$ for both walls.  Therefore, the ratios in
Eq.(\ref{e21}) should go to unity. The numerical results in Table
\ref{tab3} confirm this expectation, leading to a good agreement
between entropy production and exponential phase-space contraction
rate.
\subsubsection{Shear Flow}
We follow the same procedure as in the preceding section. For a
stationary shear flow the hydrodynamic entropy production $\sigma$ per
unit volume in Eq.\ (\ref{e18}) can be written as \cite{ChLe95,EvMo90}
\begin{equation}\label{e22} 
\sigma(y)=\frac{\Pi}{T}\frac{du_x}{dy}+J(y)\frac{d}{dy}\left(\frac{1}{T}\right) =\Pi\frac{d}{dy}\left(\frac{u_x}{T}\right).
\end{equation}
The second step in Eq.(\ref{e22}) follows from the fact that in NSS
$\lambda dT/dy=J(y)=\Pi u_x(y)$. The total  entropy
production $\overline R$ in the steady shear flow state is then
\cite{ChLe95,ChLe97} 
\begin{eqnarray}\label{e23}
\overline R = \int_{Volume} \!\!\!\sigma d{\bf r}=\int_{Surface}\!\!\!\Pi u/T ds\nonumber\\
= J_w/T_w=2L^2\Pi\left( u_w/L\right)/T_w=L^2\Pi\gamma/T_w.
\end{eqnarray}
 In the macroscopic formulation of irreversible
thermodynamics Eq.(\ref{e23}) is interpreted as an equality, in the
stationary state, between the  entropy produced in the
interior and the entropy flow carried across the
walls. Our shift map ${\cal S}_d$ mimics moving walls with drift
velocities $\pm u_w$. The work performed at these walls is converted
by the viscous bulk into heat and then again absorbed by the walls
which now act as infinite thermal reservoirs. By imagining that the
walls act as an 'equilibrium' thermal bath at temperature $T_w$, $\overline
R$ can be interpreted as their entropy increase rate.

\begin{figure}
\begin{center}
\epsfxsize=6.5cm
\subfigure[]{\epsfbox{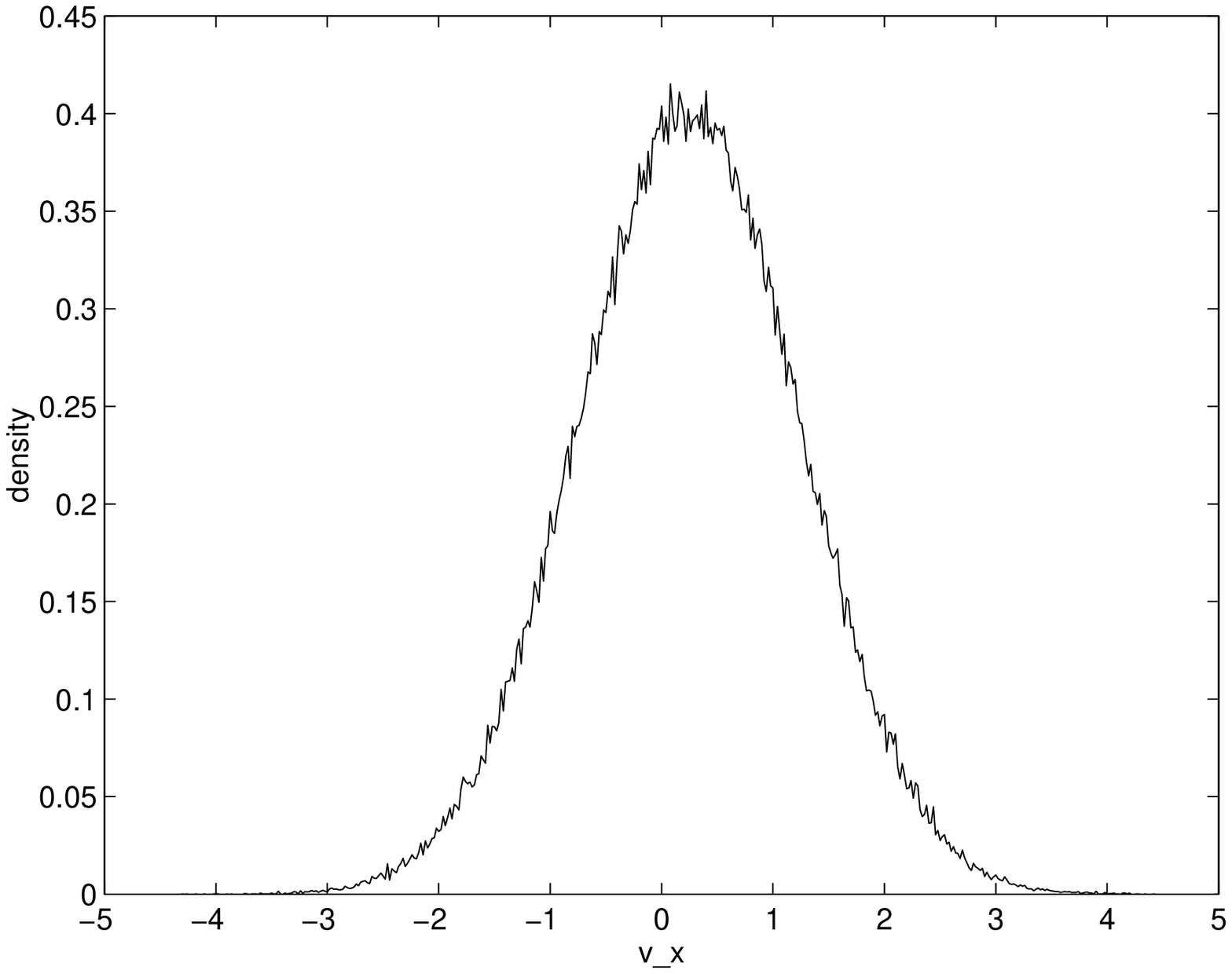}}
\epsfxsize=6.5cm
\subfigure[]{\epsfbox{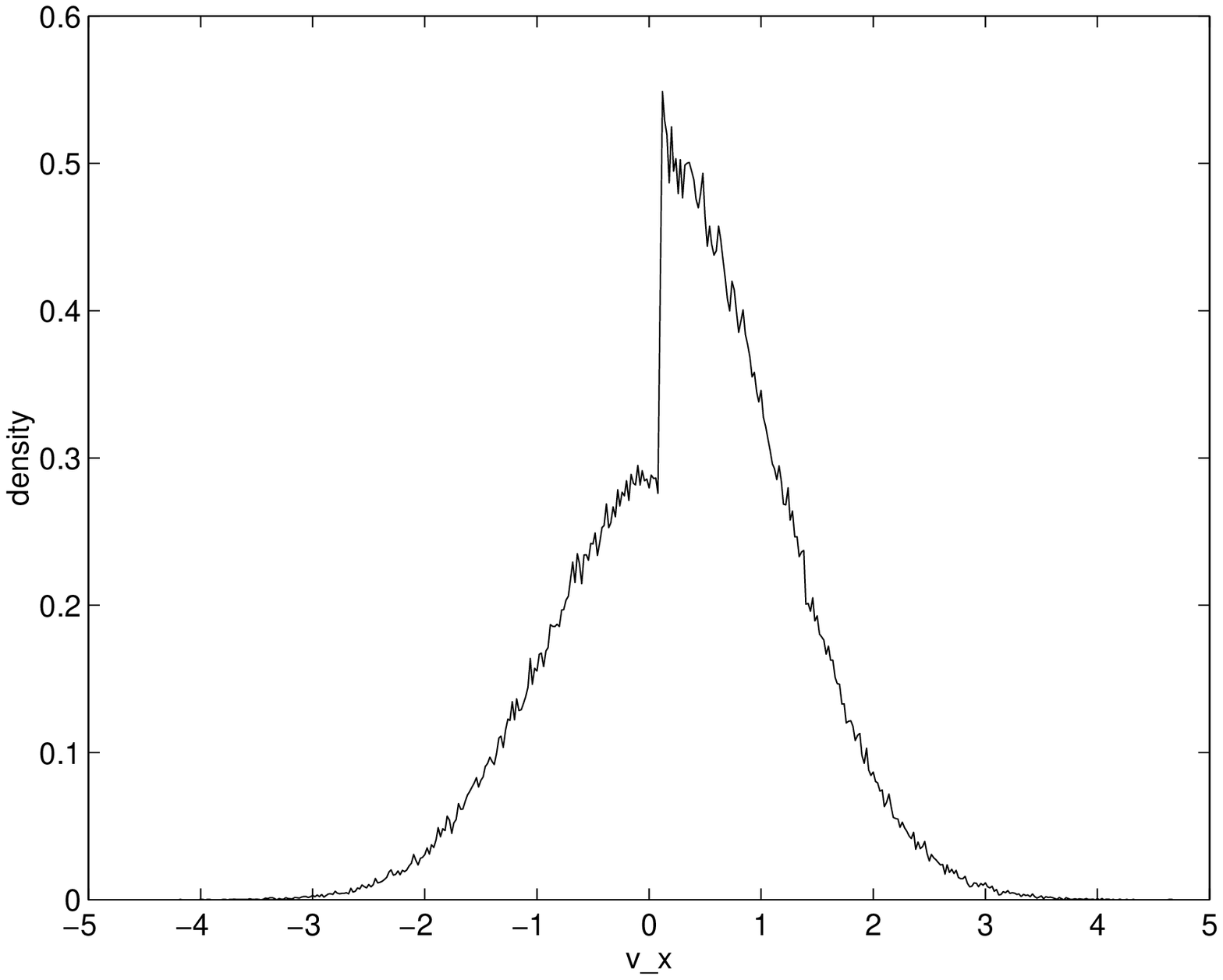}}\\
\epsfxsize=6.5cm
\subfigure[]{\epsfbox{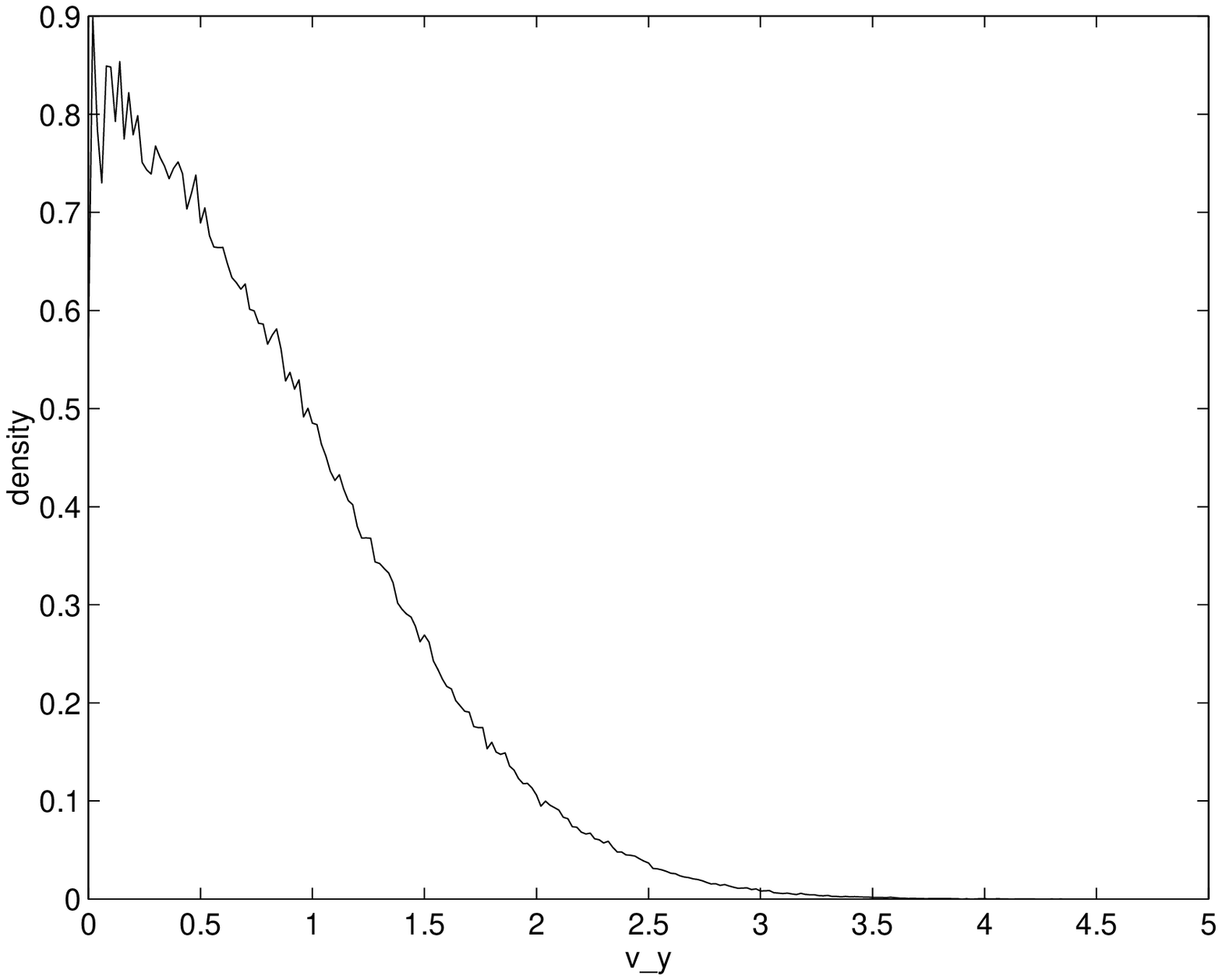}}
\epsfxsize=6.5cm
\subfigure[]{\epsfbox{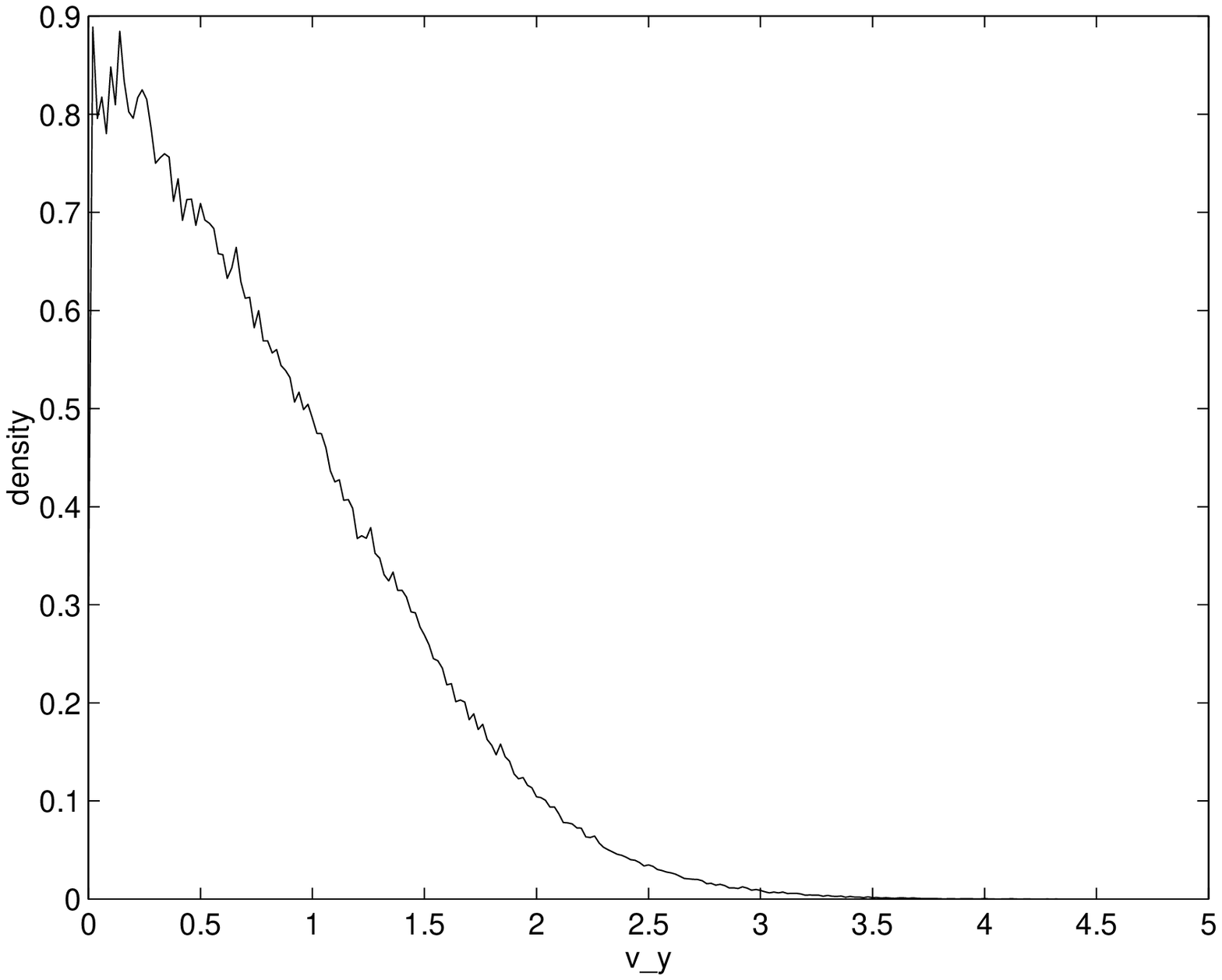}}
\end{center}
\caption{Velocity distributions of the in- and outgoing
particles at the upper wall for the shear flow case (Model I, standard
map): a) $v_x^{in}$, b) $v_x^{out}$, c) $v_y^{in}$, d) $v_y^{out}$.}
\label{fig7}
\end{figure}

\begin{figure}
\begin{center}
\epsfxsize=6.5cm
\subfigure[]{\epsfbox{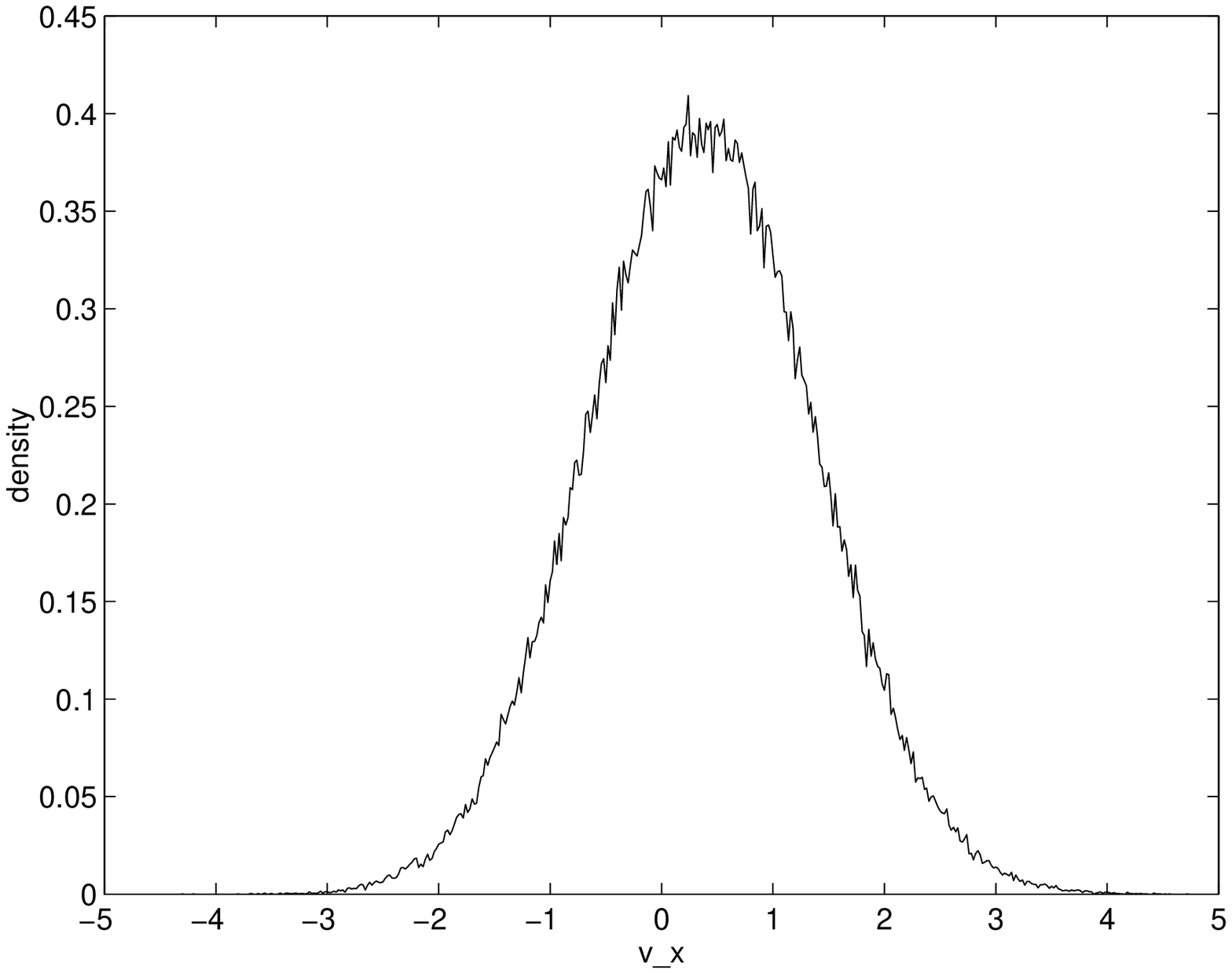}}
\epsfxsize=6.5cm
\subfigure[]{\epsfbox{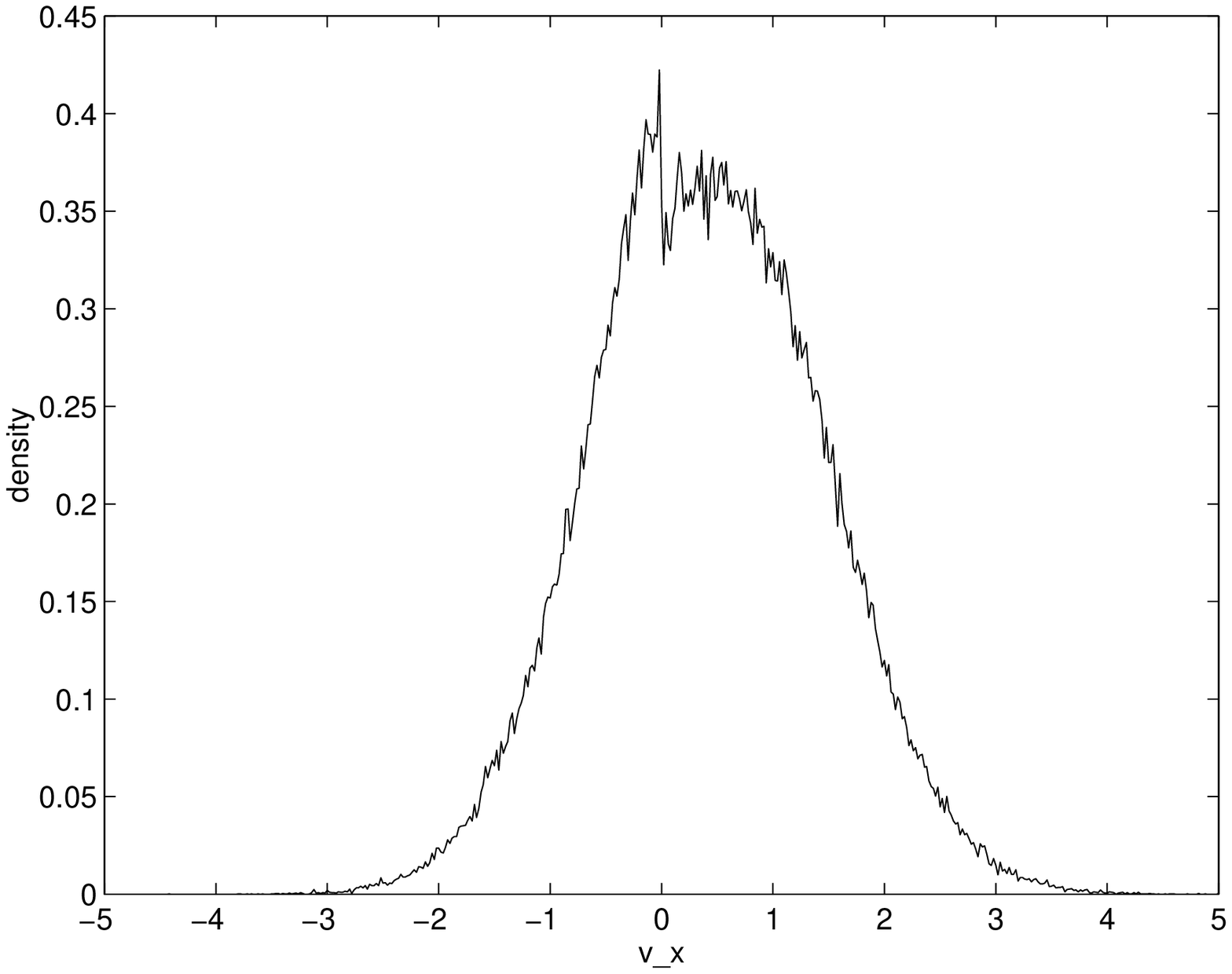}}
\end{center}
\caption{Velocity distributions of the in- and outgoing
particles at the upper wall for the shear flow case (Model II): a)
$v_x^{in}$, b) $v_x^{out}$.}
\label{fig8}
\end{figure}

For Model I (Eqs.(\ref{e11})(\ref{e11b})) the mean exponential phase-space
contraction rate takes the form ($\partial {\cal S}_d/\partial v_x \partial v_y\equiv
1$)
\begin{equation} \label{e24}
\overline P=-\left<[v_x'^2+v_y'^2-v_x^2-v_y^2-2d(v_x'+v_x)]/2T\right> ,
\end{equation}
whereas the entropy production  is  given by
\begin{equation} \label{e25}
J_w/T_w=-\left<[v_x'^2+v_y'^2-v_x^2-v_y^2-<v_x'>^2+<v_x>^2]/2T_w\right> .
\end{equation}
In Table \ref{tab4} the ratios of $L^2\Pi\gamma/J_w$ as obtained from
the simulations are reported and the relation of phase-space
contraction rate to entropy production is subsequently
checked. Whereas the equality between entropy production and entropy
flow via heat transfer is confirmed, we observe a significant
difference between entropy production and phase-space contraction
which subsists in the hydrodynamic limit. This mismatch is perhaps not
so unexpected in view of the distorted outgoing fluxes (see
Fig.\ref{fig6}).  Nevertheless, one could argue that the fine
structure of these distributions may depend on the specific
characteristics of the baker map chosen to model the collision
process. We therefore also considered the standard map
\begin{equation}\label{e26}
\widetilde {\cal M}:\left\{\begin{array}{l}\xi'=\xi-\frac{k}{2\pi}\sin(2\pi \zeta),\\
\zeta'=\zeta+\xi',\end{array}\right.
\end{equation}
with the parameter $k=100$ to ensure that we are in the hyperbolic
regime \cite{Ott,Meis92}. We found that the discrepancy
between entropy production and phase-space contraction rate
remains: although the deterministic map now seems to be chaotic enough
to smooth out the fine structure of the outgoing densities, the
discontinuity at $d$ survives. Actually, as long as  Model I is adopted it becomes clear that in
NSS there will always be more out- than ingoing particles with
$v_x\ge d$ at the upper wall (and with $v_x\le d$ at the lower wall).
Thus, the Gaussian halves in Fig.\ref{fig7} (b) will never match to a
full Gaussian even in the hydrodynamic limit, and $\Phi_i$ and
$\Phi_o$ will never come close to a local equilibrium. To circumvent
this problem we modify the map $\mbox{\boldmath${\cal T}$ }$ and investigate the following
model:
\begin{eqnarray}
\mbox{(Model II)}&&\nonumber\\
\mbox{\boldmath${\cal T}$ }\!\!_{\pm}(v_x,v_y)&=&\left(\frac{\mbox{erf}\left[(|v_x|\mp
d)/\sqrt{2T}\right]\pm\mbox{erf}(d/\sqrt{2T})}{ 1\pm
\mbox{erf}(d/\sqrt{2T})} \, , \exp(-v_y^2/2T) \right)\label{e27}\\ 
\nonumber\\
(v_x',v_y')&=&\mbox{\boldmath${\cal T}$ }\!\!_+^{-1}\circ\widetilde{\cal M}\circ\mbox{\boldmath${\cal T}$ }\!\!_- (v_x,v_y) ,
\qquad v_x\ge 0 \label{e27a}\\ 
(v_x',v_y')&=&-\mbox{\boldmath${\cal T}$ }\!\!_-^{-1}\circ\widetilde{\cal
M}^{-1}\circ\mbox{\boldmath${\cal T}$ }\!\!_+ (v_x,v_y) , \quad v_x<0\nonumber .
\end{eqnarray}     
This model is also time-reversible, but in contrast to the former one no
particle changes its tangential direction during the scattering.
There is still a gap in the outgoing distribution of Fig.(\ref{fig8})
(b), however, simulations show that this gap disappears in the
hydrodynamic limit thus bringing the in- and the outgoing
distributions close to local equilibrium. Furthermore, we note that
whereas we were not able to give a relation between the parameter $d$
and the actual wall velocity $u_w$ for Model I, in case of Model II
$u_w$ converges to $d$ in the hydrodynamic limit. For this reason we
chose $d=0.5$ in the following, since this value yields the same order
of the wall velocity as $d=0.1$ for Model I. Proceeding now to the
phase-space contraction rate we find that it takes the form
\begin{equation} \label{e28}
\overline P^{u/d}=-(n_+-n_-) \ln\frac{1\pm \mbox{erf}(d/\sqrt{2T})}{1\mp
\mbox{erf}(d/\sqrt{2T})}  -
\left<[v_x'^2+v_y'^2-v_x^2-v_y^2-2d(v_x'+v_x)]/2T\right>_{u/d}\: ,
\end{equation}
where $n_{\pm}$ are the collision rates for positive and negative tangential
velocities, and the additional term (cf. Eq.(\ref{e24})) results from
the different denominators in Eq.(\ref{e27}). Note that one has to
average over the upper and the lower wall separately. Again we compare
$\overline R$ and $\overline P$ (Table \ref{tab5}), but although the outgoing flux
approaches now a Gaussian in the hydrodynamic limit the two quantities still do
not match. This result can be understood in more detail by
 rearranging the terms in Eq.(\ref{e28}) as
\begin{eqnarray} \label{e28a}
\overline
P^{u/d}&=&-\left<[v_x'^2+v_y'^2-v_x^2-v_y^2-<v_x'>^2+<v_x>^2]/2T\right>_{u/d}\nonumber\\ 
&&-\left<[<v_x'>^2-<v_x>^2-2d(v_x'+v_x)]/2T\right>_{u/d}
-(n_+-n_-) \ln\frac{1\pm \mbox{erf}(d/\sqrt{2T})}{1\mp
\mbox{erf}(d/\sqrt{2T})}\: .
\end{eqnarray}
Since $T\to T_w$ in the hydrodynamic limit, the first term clearly
corresponds to the entropy production Eq.(\ref{e25}). However, the
second and the third terms provide additional contributions.  For
$u_w\to d$ and $d\to 0$ they are both of order $d^2$ and can be
interpreted as a phase space contraction due to a friction {\em
parallel} to the walls \cite{NiDa98}. These two terms apparently
depend on the specific modeling of the collision process at the
wall. They may physically be interpreted as representing certain
properties of a wall, like a roughness, or an anisotropy. Actually, the
second term already appeared in Model I, see Eq.(\ref{e24}). The price
we had to pay in Model II for the fluxes getting close to local
equilibrium is the additional third term in Eq.(\ref{e28a}), which
does not compensate the second one.

\begin{figure}
\begin{center}
\epsfxsize=6.5cm
\subfigure[]{\epsfbox{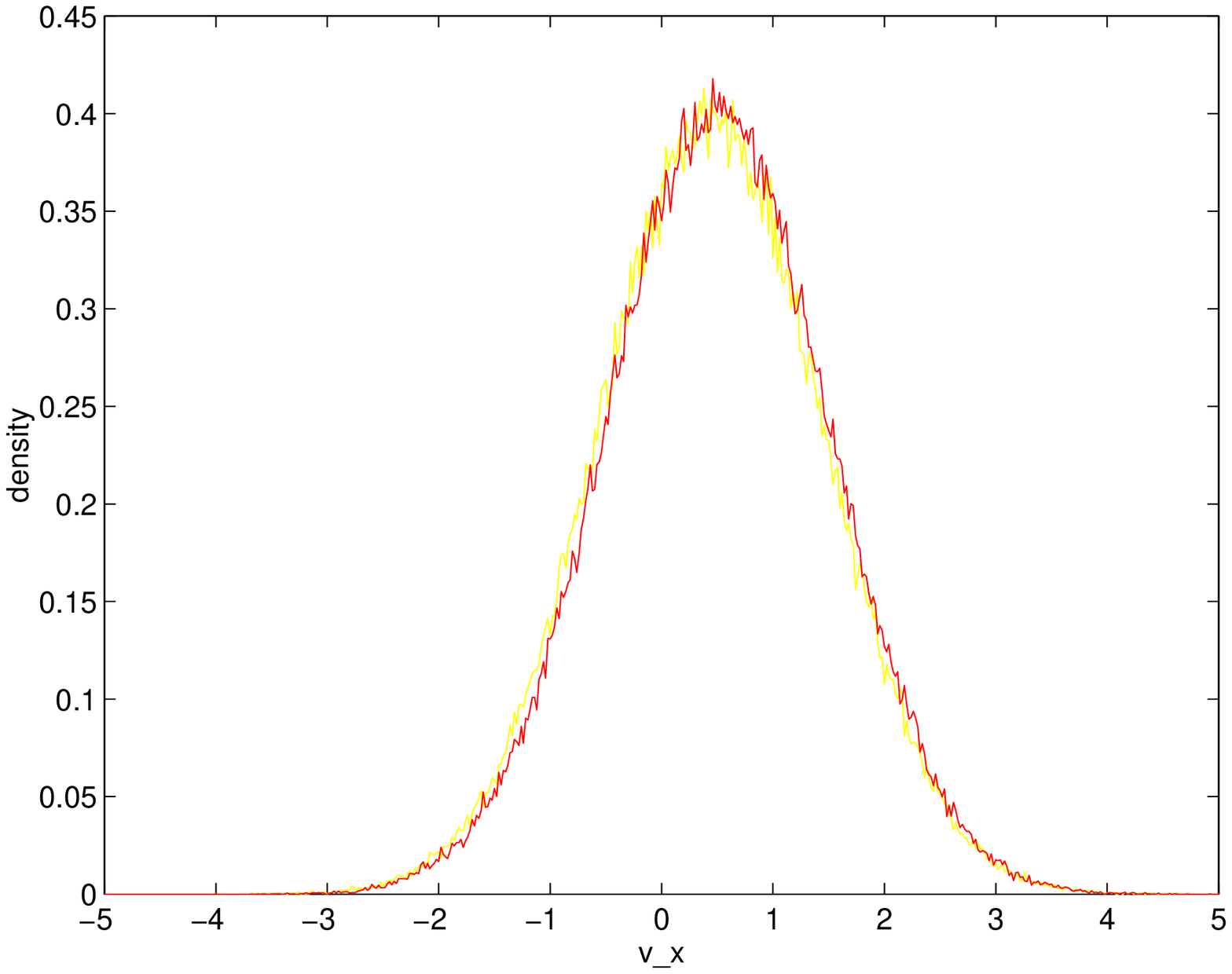}}
\epsfxsize=6.5cm
\subfigure[]{\epsfbox{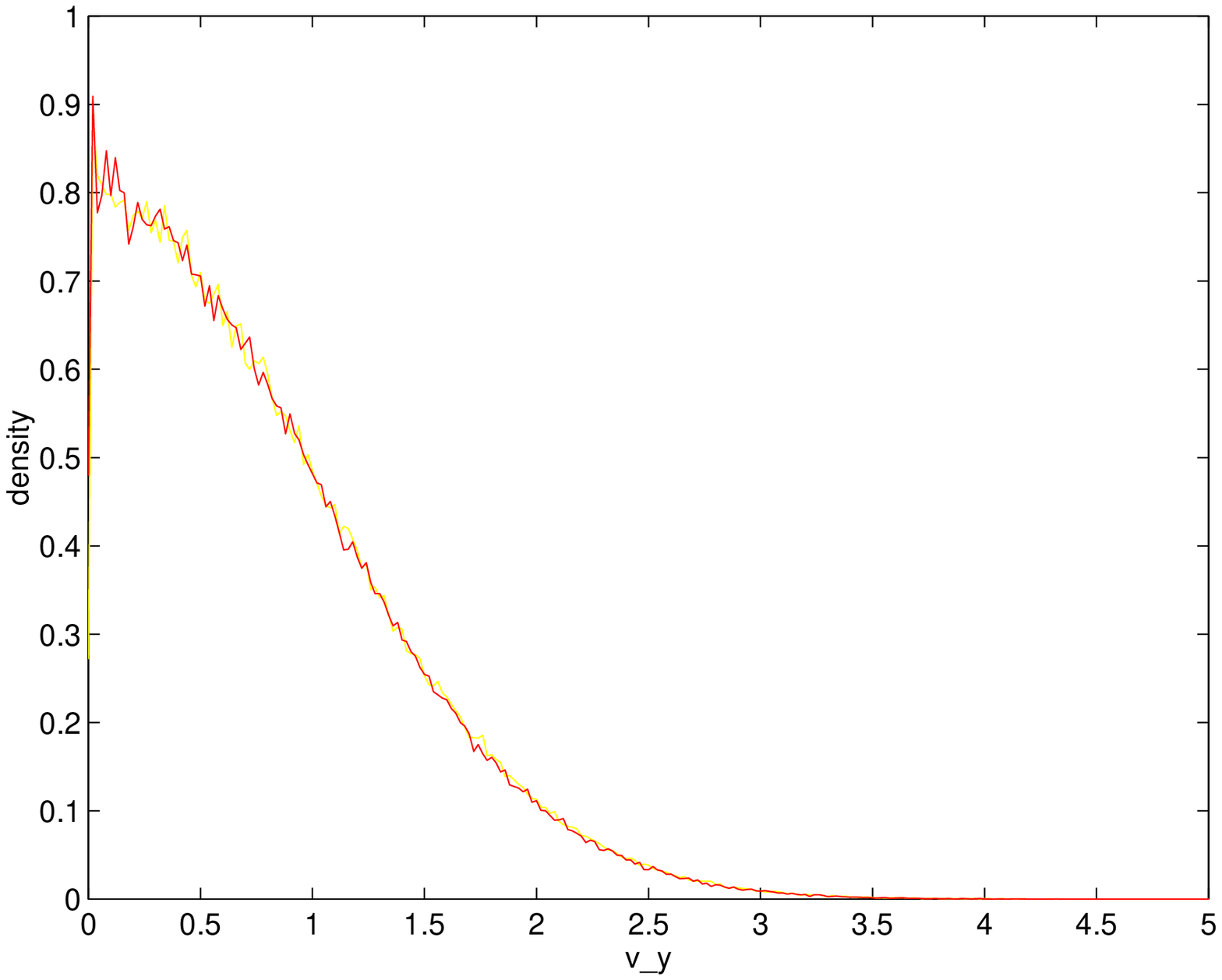}}
\end{center}
\caption{Velocity distributions of the in- and outgoing
particles at the upper wall for the shear flow case (Model III):  a)
$v_x^{in},v_x^{out}$, b) $v_y^{in},v_y^{out}$.}
\label{fig9}
\end{figure}

The foregoing analysis shows clearly what to do to get rid of the
additional term in Eq.(\ref{e28}): we have to use the same forward and
backward transformations $\mbox{\boldmath${\cal T}$ }_{\pm}$ in
Eqs.(\ref{e27},\ref{e27a}).  If one still wants to transform onto a
full Gaussian in the hydrodynamic limit time-reversibility has to be
given up. This leads us to propose the model
\begin{eqnarray}\label{e29}
\mbox{(Model III)}&&\nonumber\\
\mbox{\boldmath${\cal T}$ }\!\!_{*}(v_x,v_y)&=&\left(\frac{\mbox{erf}\left[(v_x-d)/\sqrt{2T}\right]+1}{2}
\, , \exp(-v_y^2/2T) \right) \\ \nonumber\\
(v_x',v_y')&=&\mbox{\boldmath${\cal T}$ }\!\!_*^{-1}\circ\widetilde{\cal M}\circ\mbox{\boldmath${\cal T}$ }\!\!_*(v_x,v_y) . 
\end{eqnarray}      
which is still deterministic, but no longer time-reversible. The
phase-space contraction is now given as
\begin{equation} \label{e30}
\overline P=   \left<[v_x'^2+v_y'^2-v_x^2-v_y^2-2d(v_x'-v_x)]/2T\right> .
\end{equation}
Fig. \ref{fig9} shows that the in- and the outgoing fluxes are getting
close to local equilibrium, implying that the velocity of the wall
$u_w$ goes to $d$ and the wall temperature $T_w$ goes to
$T$. Consequently, Eq.(\ref{e30}) should converge to the correct
thermodynamic entropy production of Eq.(\ref{e25}) in the hydrodynamic
limit, and this is indeed what we observe in Table \ref{tab5}. This
implies that time-reversibility does not appear to be an essential
ingredient for having a relation between phase space contraction and
entropy production, as was already stated in Refs.\
\cite{ChLe95,ChLe97,NiDa98}. We remark that we consider the lack of
time-reversibility in Model III rather as a technical difficulty of
how we define our scattering rules than a fundamental property of this
model.

\section{Conclusion}\label{sec4}
We have  applied a novel thermostating mechanism to an
interacting many-particle system. Under this formalism the system is
thermalized through scattering at the boundaries while the bulk is
left Hamiltonian. We have shown how this deterministic and
time-reversible thermostating mechanism is related to conventional
stochastic boundary conditions. For a two-dimensional system of hard
disks, this thermostat yields a stationary nonequilibrium heat or
shear flow state. Transport coefficients obtained from computer
simulations, such as thermal conductivity and viscosity, agree with
the values obtained from Enskog's theory.

Having a time-reversible and deterministic system we also examined the
relation between microscopic reversibility and macroscopic
irreversibility in terms of entropy production. We find that
 entropy production and exponential phase-space
contraction rate do in general not agree. When the NSS is created by a temperature
gradient both
quantities converge in the hydrodynamic limit. By subjecting the
system to a shear we examined three different versions of scattering
rules, of which one (Model III) produced  an agreement.

Our results indicate that neither time-reversibility nor the existence
of a local thermodynamic equilibrium at the walls are sufficient
conditions for obtaining an identity between phase space contraction
and entropy production. A class of systems where such an identity is
guaranteed by default are the ones thermostated by velocity-dependent
friction coefficients \cite{EvMo90,Hoov91,MoDe98}.  We suggest that in
general, that is, by using other ways of deterministic and
time-reversible thermostating, such an identity may not necessarily
exist. We would expect the same to hold for any system where the
interaction between bulk and reservoir depends on the details of the
microscopic scattering rules.

As a next step it would be important to compute the spectrum of
Lyapunov exponents for the models presented in this paper. This would
enable to check, for example, the validity of formulas which express
transport coefficients in terms of sums of Lyapunov exponents
\cite{PoHo88,ECM}, and the existence of a so-called conjugate
pairing rule of Lyapunov exponents \cite{ECM,DePo97b}. Moreover, it
would be interesting to verify the fluctuation theorem, as it has been
done recently for the Chernov-Lebowitz model \cite{BCL98}.

Acknowledgments\\ Helpful discussions with P.Gaspard, M.Mareschal and
K. Rateitschak are gratefully acknowledged. R.K. wants to thank the
Deutsche Forschungsgemeinschaft (DFG) for financial support. This work
is supported, in part, by the Interuniversity Attraction Pole program
of the Belgian Federal Office of Scientific, Technical and Cultural
Affairs and by the Training and Mobility Program of the European
Commission.


\begin{table}
\caption{\label{tab1} \bf Comparison of theoretical and experimental heat conductivity $\lambda_{exp}/\lambda_{th}$ }
\begin{tabular}{lllll}
&N=100&N=200&N=400&N=800 \\ \hline $\Delta T$=1.5-1
&0.904&1.009&1.003&1.062\\ $\Delta T$=2.0-1 &0.887&0.950&1.021&1.051
\end{tabular}
\end{table}

\begin{table}
\caption{ \label{tab2}\bf Comparison of theoretical and experimental viscosity $\eta_{exp}/\eta_{th}$ } 
\begin{tabular}{lllll}
&N=100&N=200&N=400&N=800 \\ \hline
$d$=0.05 &0.9616&0.9904&1.0081&1.0382\\
$d$=0.1 &0.9702&1.001&1.0226&1.0232
\end{tabular}
\end{table}

\begin{table}
\caption{ \label{tab3}\bf Comparison of entropy production and exp. phase-space contraction rate $\overline R/\overline P$, heat flow} 
\begin{tabular}{lllll}
&N=100&N=200&N=400&N=800 \\ \hline
$T_u=1.5$&1.0814&1.0762&1.0614&1.0508\\
$T_d=1$&0.8948&0.9170&0.9273&0.9439\\ \hline
$T_u=2$&1.1313&1.1110&1.0985&1.0765\\
$T_d=1$&0.8122&0.8412&0.8633&0.8886
\end{tabular}
\end{table}

\begin{table}
\caption{ \label{tab4}\bf Comparison of entropy production and exp. phase-space contraction rate $\overline R/\overline P$, shear flow, Model I} 
\begin{tabular}{clllll}
&&N=100&N=200&N=400&N=800 \\ \hline
$d$=0.05&$L^2\Pi\gamma/J_w$&0.9816&0.9669&0.9765&0.9962\\
baker map&$\overline R/\overline P$&0.6761&0.5882&0.5023&0.4230\\ \hline
$d$=0.1&$L^2\Pi\gamma/J_w$&0.9829&0.9664&0.9497&0.9588\\
baker map&$\overline R/\overline P$&0.6457&0.5761&0.4934&0.4275\\ \hline
$d$=0.1&$L^2\Pi\gamma/J_w$&1.0255&1.0424&1.0435&1.0833\\
standard map&$\overline R/\overline P$&0.4622&0.3873&0.3008&0.2417
\end{tabular}
\end{table}

\begin{table}
\caption{ \label{tab5}\bf Comparison of entropy production and exp. phase-space contraction rate $\overline R/\overline P$, shear flow, Model II and III} 
\begin{tabular}{clllll}
&&N=100&N=200&N=400&N=800 \\ \hline
Model II&$L^2\Pi\gamma/J_w$&0.9842&1.0154&1.0002&1.0255\\
$d$=0.5&$\overline R/\overline P$&0.1858&0.1682&0.1398&0.1127\\ \hline
Model III&$L^2\Pi\gamma/J_w$&1.0164&1.0046&1.0005&1.0039\\
$d$=0.5&$\overline R/\overline P$&0.8452&0.8785&0.9051&0.9619
\end{tabular}
\end{table}


\begin{thebibliography}{10}
\bibitem{alter}
We note that in our numerical experiments we alternate the application of
  ${\cal M}$ and ${\cal M}^{-1}$ in Eq.(\ref{e4}) with respect to the position
  of the scattering process at the wall to avoid any artificial symmetry
  breaking.
\end{thebibliography}
\end{document}